\documentstyle[11pt]{article}

\newfont{\ssr}{cmss10 scaled 1100}

\makeatletter
\@addtoreset{equation}{section}
\makeatother

\def\dalemb#1#2{{\vbox{\hrule height .#2pt
        \hbox{\vrule width.#2pt height#1pt \kern#1pt
                \vrule width.#2pt}
        \hrule height.#2pt}}}

\let\a=\alpha    \let\e=\epsilon

    \let\c=\chi

 \def\bd{\begin{document}} \def\ed{\end{document}}
\def\ds{\documentstyle} \let\fr=\frac \let\bl=\bigl \let\br=\bigr
\let\Br=\Bigr \let\Bl=\Bigl 
\let\bm=\bibitem
\let\na=\nabla
\let\pa=\partial \let\ov=\overline 
\newcommand{\be}{\begin{equation}} 
\newcommand{\ee}{\end{equation}} 
\def\ba{\begin{array}}
\def\ea{\end{array}}
\def\ft#1#2{{\textstyle{{\scriptstyle #1}\over {\scriptstyle #2}}}}
\def\fft#1#2{{#1 \over #2}}
\def\del{\partial}
\def\sst#1{{\scriptscriptstyle #1}}
\def\oneone{\rlap{\rm 1}\mkern4mu{\rm l}}
\def\e7{E_{7(+7)}}
\def\td{\tilde}
\def\bog{Bogomol'nyi\ }
\def\Z{\rlap{\ssr Z}\mkern3mu\hbox{\ssr Z}}
\def\R{\rlap{\rm I}\mkern3mu{\rm R}}
\newcommand{\ho}[1]{$\, ^{#1}$}
\newcommand{\hoch}[1]{$\, ^{#1}$}
\newcommand{\bea}{\begin{eqnarray}} 
\newcommand{\eea}{\end{eqnarray}} 
\newcommand{\ra}{\rightarrow}
\newcommand{\lra}{\longrightarrow}
\newcommand{\Lra}{\Leftrightarrow}
\newcommand{\ap}{\alpha^\prime}
\newcommand{\bp}{\tilde \beta^\prime}
\newcommand{\tr}{{\rm tr} }
\newcommand{\Tr}{{\rm Tr} } 
\newcommand{\NP}{Nucl. Phys. }

\begin{document}
\begin{flushright}
\hfill{CTP-TAMU-7/96}\\
\hfill{Imperial/TP/95--96/30}\\
\hfill{hep-th/9602140}\\
\end{flushright}

\vspace{20pt}

\begin{center} { \large {\bf Weyl Group Invariance and $p$-brane
Multiplets}}

\vspace{30pt}

H. L\"u\hoch{\dagger}, C.N. Pope\hoch{\dagger}

\vspace{15pt}

{\it Center for Theoretical Physics, Texas A\&M University, College
Station, Texas 77843}

\vspace{10pt}

K.S. Stelle\hoch{\sst\star}

\vspace{15pt}

{\it The Blackett Laboratory, Imperial College, Prince Consort Road,
London SW7 2BZ, UK} 

\vspace{40pt}

\underline{ABSTRACT}
\end{center}

      In this paper, we study the actions of the Weyl groups of the U
duality groups for type IIA string theory toroidally compactified to all
dimensions $D\ge 3$.  We show how these Weyl groups implement permutations
of the field  strengths, and we discuss the Weyl group multiplets of all
supersymmetric $p$-brane solitons. 

{\vfill\leftline{}\vfill
\vskip	10pt
\footnoterule {\footnotesize
	\hoch{\dagger}	Research supported in part by DOE  Grant
DE-FG05-91-ER40633 \vskip	-12pt}  \vskip	10pt {\footnotesize 
        \hoch{\sst\star} Research supported in part by the Commission
of the  European Communities under contract SCI*-CT92-0789} }

\pagebreak
\setcounter{page}{1}

\section{Introduction}

      It has been known for a long time that the maximal supergravity
theory in $D$ dimensions obtained by Kaluza-Klein dimensional reduction
from $D=11$ has a $G\simeq E_{n(+n)}(\R)$ symmetry, where $n=11-D$
\cite{cj}. Recently it was conjectured that the $G(\Z)\simeq
E_{n(+n)}(\Z)$ discrete subgroup of this is an exact symmetry of the
corresponding string theory, known as U duality \cite{dl,ht}.  In 
general, the solutions of a supergravity theory will form
infinite-dimensional multiplets under U duality.   In \cite{ht}, this
symmetry was  used to fill out a full set of 56 purely electric and
purely magnetic 4-dimensional black holes that break half of the
supersymmetry of $N=8$ supergravity. Of course, these are only a finite
subset of the infinite dimensional $E_{7(+7)}(\Z)$ multiplet; the other
members of the multiplet have more complicated field-strength
configurations involving both electric and magnetic contributions,
together with some non-vanishing axions ({\it i.e.}\ 0-form potentials
for 1-form field strengths).  However, since the 56 purely electric and
purely magnetic solutions are of primary interest, in view of their
interpretation as fundamental quantum states of the string theory, it is
useful to isolate a subgroup of $E_{7(+7)}(\Z)$ that maps these solutions
among themselves. We shall show that the group that does this is the Weyl
group of $E_{7(+7)}$.  In this paper, we shall study the U Weyl group
invariance  of all maximal supergravity theories in dimensions $D\ge 3$.  We
shall also study the corresponding multiplet structures of $p$-brane
solutions under the U Weyl group. 

     The significance of the Weyl subgroup $W$ of the U duality group $G$ is
analogous to that of the $\Z_2$ subgroup of the $U(1)$ electric-magnetic
duality group in Maxwell theory, describing the discrete interchange of
electric and magnetic fields: $E\longrightarrow B$ and $B\longrightarrow
-E$. Thus $W$ is the subgroup of the U duality group that implements certain
permutations of the field strengths in the theory, while maintaining their
alignment along the axes of the space of field strengths.  In other words,
the U Weyl group describes certain 90-degree rotations in the space of field
strengths. By contrast, the full U duality group $G$, as well as its quantum
restriction $G(\Z)$, include  intermediate rotations in the space of field
strengths. In terms of $p$-brane solutions, therefore, the U Weyl group
preserves the total number of electric and magnetic charges, whereas the
full U duality group does not. Thus the U Weyl group gives a
characterisation of the independent $p$-brane solutions of a given type. 

     The U Weyl group $W$ has another interpretation when acting on the
space of supergravity solutions. The scalar fields of a supergravity
theory admitting flat space as a solution ({\it i.e.}\ of an
``ungauged'' supergravity) do not have potentials, and hence their
asymptotic values are unfixed by the equations of motion. These
asymptotic values constitute the moduli of a supergravity solution;
since certain subsets of the scalars also occur in exponential
prefactors of the antisymmetric tensor field kinetic terms, the moduli
also determine the coupling constants of the theory in a particular
asymptotically-defined ``vacuum.'' Performing a $G(\Z)$ U duality
transformation on a supergravity solution will generally change the
asymptotic values of the scalars, and hence change the ``vacuum.'' A
subgroup of $G(\Z)$ exists, however, that does not change the vacuum.
This stability duality group of the vacuum coincides precisely with the
Weyl group as defined above when the scalars are taken to tend
asymptotically to zero, which will be the case for the family of
$p$-brane solutions considered here.

     In identifying the subgroup $W$ of a U duality group $G$ that
preserves the total number of electric and magnetic charges as the Weyl
group of $G$, we are not using the classic definition \cite{helg} of
the Weyl group as the quotient of the normalizer divided by the
centralizer of the Cartan subalgebra of $G$. Since this standard
definition is a quotient-group construction, it is not even clear in
general that its result properly defines a subgroup of $G$.
Nonetheless, the group $G$ may still have discrete subgroups that are
isomorphic to $W$, even though they are not obtained by the classic
quotient construction. This will be the case for the discrete subgroups
considered in this paper. We shall identify them as Weyl groups by
their explicit actions on the vectors of scalar fields (``dilatonic
scalars'') appearing in the exponential prefactors of antisymmetric
tensor field strengths in the action. These dilaton vectors will
transform under the action of our discrete $W$ subgroups in exactly the
same way as weight vectors of the irreducible representations of $G$
transform under the Weyl groups, thus giving us the identification.

     The U duality group $G(\Z)$ has several important subgroups,
namely the T, S, and X dualities. T duality \cite{gpr} is a
perturbative symmetry of string theory, valid order by order in the
string coupling constant $g$, although non-perturbative in $\a'$.  It
is a $D_{n-1} \simeq SO(n-1, n-1;\Z)$ subgroup of the U duality group
$G(\Z)\simeq E_{n(+n)}(\Z)$, which acts on the compactified internal
dimensions. The T duality group preserves the NS-NS and R-R sectors of
the theory. X duality is an $SL(2,\Z)$ subgroup of $E_{n(+n)}(\Z)$ that
interchanges the NS-NS with the R-R fields. It is a conjectured
non-perturbative symmetry of the string theory. In the type IIB
context, it is already present in $D=10$ \cite{sch1}; in type IIA, it
is present for
$D=9$ \cite{bho} and lower. Finally, S duality \cite{dl}, which exists
only in $D=4$, is another $SL(2,\Z)$ subgroup of $E_{7(+7)}(\Z)$, again
non-perturbative. One way to see why four is a special dimension in this
regard is to note that whereas the T duality group $D_{n-1}$ is a
maximal subgroup of $E_n$ for general values of $n$, the case $n=7$ is
special, since then $D_6\times A_1$ is a maximal subgroup of $E_7$. 
The $A_1$ factor corresponds precisely to the $SL(2,\Z)$ S duality.   S
duality  preserves the NS-NS and R-R sectors; in particular, it rotates
between each NS-NS 2-form field strength and its dual \cite{filq}.  To
summarise, both T and S duality preserve the NS-NS and R-R sectors, the
former being  perturbative whilst the latter is non-perturbative,
interchanging strong and weak couplings. X duality is a
non-perturbative symmetry, which interchanges the NS-NS and R-R
sectors.  In this paper, we shall also discuss the Weyl group
symmetries of the S, T and X duality groups. As in the case of the U
Weyl group, the S, T and X Weyl groups also describe certain 90-degree
rotations in the space of field strengths. 

     The bosonic sector of the Lagrangian for $D=11$ supergravity is 
${\cal L}= \hat e \hat R -\ft1{48} \hat e\, \hat F_4^2 +\ft16 
\hat F_4\wedge\hat F_4\wedge \hat A_3$, where $\hat A_3$ is the 3-form
potential for the 4-form field strength $\hat F_4$.  Upon Kaluza-Klein
dimensional reduction to $D$ dimensions, this yields the following 
Lagrangian:
\bea {\cal L} &=& eR -\ft12 e\, (\del\vec\phi)^2 -\ft1{48}e\, e^{\vec
a\cdot 
\vec\phi}\, F_4^2 -\ft{1}{12} e\sum_i  e^{\vec a_i\cdot \vec\phi}\,
(F_3^{i})^2 -\ft14 e\, \sum_{i<j} e^{\vec a_{ij}\cdot \vec\phi}\,
(F_2^{ij})^2
\nonumber\\ && -\ft14e\, \sum_i e^{\vec b_i\cdot \vec\phi}\, ({\cal
F}_2^i)^2 -\ft12 e\, \sum_{i<j<k} e^{\vec a_{ijk} \cdot\vec \phi}\,
(F_1^{ijk})^2 -\ft12e\, \sum_{i<j} e^{\vec b_{ij}\cdot \vec\phi}\,
({\cal F}_1^{ij})^2 + {\cal L}_{\sst{FFA}}\ ,\label{dgenlag}
\eea where $F_4$, $F_3^i$, $F_2^{ij}$ and $F_1^{ijk}$ are the 4-form,
3-forms, 2-forms and 1-forms coming from the dimensional reduction of
$\hat F_4$ in
$D=11$; ${\cal F}_2^i$ are the 2-forms coming from the dimensional
reduction of the vielbein, and ${\cal F}_1^{ij}$ are the 1-forms coming
from the dimensional reduction of these 2-forms.  The quantity
$\vec\phi$ denotes an
$(11-D)$-component vector of scalar fields, which we refer to as
dilatonic scalars, arising from the dimensional reduction of the
elfbein.  These scalars appear undifferentiated, {\it via} the
exponential prefactors for the antisymmetric-tensor kinetic terms, and
they should be distinguished from the remaining spin-0 quantities in
$D$ dimensions, namely the axions, {\it i.e.,} the 0-form potentials
$A_0^{ijk}$ and ${\cal A}_0^{ij}$.  These axion fields have constant
shift symmetries, under which the action is invariant, and are thus
properly thought of as 0-form potentials rather than true scalars.  In
particular, their 1-form field strengths can adopt topologically
non-trivial configurations, corresponding to electrically-charged or
magnetically-charged $p$-brane solitons. 

     The term ${\cal L}_{\sst{FFA}}$ comes from the dimensional
reduction of  the $\hat F_4\wedge \hat F_4 \wedge \hat A_3$ term in
$D=11$. Note that the field  strengths appearing in the kinetic terms
in (\ref{dgenlag}) in general  acquire ``Chern-Simons'' modifications,
as a consequence of the dimensional reduction procedure.  As we shall
see later in detail, both of these additional types of term play only a
subsidiary role in the discussion of the Weyl group symmetries. 

     The remaining aspect of the bosonic Lagrangian (\ref{dgenlag})
that we need to address is the set of constant vectors $\vec a_{i\cdots
j}$ and $\vec b_{i\cdots j}$ appearing in the exponential prefactors of
the kinetic terms for the various antisymmetric tensors.  As was shown
in \cite{lp1}, these ``dilatonic vectors'' can be expressed  as
follows: 
\bea &&F_{\sst{MNPQ}}\qquad\qquad\qquad\qquad\qquad\qquad {\rm
Vielbein}\nonumber\\ {\rm 4-form:}&&\vec a = -\vec g\ ,\nonumber\\ {\rm
3-forms:}&&\vec a_i = \vec f_i -\vec g \ ,\label{dilatonvec}\\ {\rm
2-forms:}&& \vec a_{ij} = \vec f_i + \vec f_j - \vec g\ ,
\qquad\qquad\qquad \,\,\, \,\vec b_i = -\vec f_i\nonumber\,\\ {\rm
1-forms:}&&\vec a_{ijk} = \vec f_i + \vec f_j + \vec f_k -\vec g
\ ,\qquad\qquad\vec b_{ij} = -\vec f_i + \vec f_j\ ,\nonumber
\eea where the vectors $\vec g$ and $\vec f_i$ have $(11-D)$ components
in $D$ dimensions, and satisfy
\be
\vec g \cdot \vec g = \ft{2(11-D)}{D-2}, \qquad
\vec g \cdot \vec f_i = \ft{6}{D-2}\ ,\qquad
\vec f_i \cdot \vec f_j = 2\delta_{ij} + \ft2{D-2}\ .\label{gfdot}
\ee Note that the definitions in (\ref{dilatonvec}) are given for
$i<j<k$, and that the vectors $\vec a_{ij}$ and $\vec a_{ijk}$ are
antisymmetric in their indices. The 1-forms ${\cal F}_{\sst{M}i}^{(j)}$
and hence the vectors
$b_{ij}$ are only defined for $i<j$, but it is sometimes convenient to
regard them as being antisymmetric too, by defining $\vec b_{ij}=-\vec
b_{ji}$ for $i>j$.  

      Eqns (\ref{dilatonvec}) and (\ref{gfdot}) contain all the
information we shall need about the dilaton vectors in $D$-dimensional
maximal supergravity.  The explicit forms of the vectors $\vec f_i$ and
$\vec g$ that result from the dimensional reduction of $D=11$
supergravity are 
\bea 
\vec g &=&3 (s_1, s_2, \ldots, s_{11-\sst D})\ ,\nonumber\\
\vec f_i &=& \Big(\underbrace{0,0,\ldots, 0}_{i-1}, (10-i) s_i, s_{i+1},
s_{i+2}, \ldots, s_{11-\sst D}\Big)\ ,\label{gfvec}
\eea where $s_i = \sqrt{2/((10-i)(9-i))}$.  Of course, the explicit
forms of these  vectors are inessential because they depend on the
specific dimensional-reduction procedure used. For example, one can
also  obtain lower dimensional supergravities by dimensional reduction
of type IIB  supergravity in $D=10$. In this case, the explicit forms
of the vectors $\vec f_i$ and $\vec g$ are different from the ones
given in (\ref{gfvec}). However, these forms are related to those of
(\ref{gfvec}) by an orthogonal transformation of the dilaton vectors
involving a rotation of the first two components,
$\phi_1$ and $\phi_2$; hence they also satisfy the dot product relations
(\ref{gfdot}).  Specifically, in the type IIB basis, the vectors are
given by 
\bea
\vec g &=& 3(0,-\fft2{3\sqrt7},,s_3,s_4,\ldots, s_{11-\sst{D}})\
,\nonumber\\
\vec f_1 &=& (1,-\fft3{\sqrt7},s_3,s_4,\ldots, s_{11-\sst{D}}) \ , 
\label{gfvecb}\\
\vec f_2 &=& (-1,-\fft3{\sqrt7},s_s,s_4,\ldots,s_{11-\sst{D}})\
,\nonumber
\eea with $\vec f_i$ for $i\ge3$ the same as in (\ref{gfvec}).

   Having obtained the full bosonic Lagrangian for $D$-dimensional
maximal  supergravity, we are in a position to discuss its symmetry
under the Weyl group of the supergravity symmetry group.  In section 2,
we shall examine the $E_{2(+2)}(\R) \simeq GL(2,\R)$ symmetry group  of
$D=9$ supergravity in detail, and show how the discrete $\Z_2$ Weyl
group emerges as the group of permutations of pairs of field-strength
tensors.  In addition, we shall show that this $\Z_2$ Weyl group
symmetry exists also in type IIB  supergravity in
$D=10$.   In section 3, we generalise the discussion to maximal
supergravities in  $3\le D\le 8$ dimensions.  In section 4, we shall
discuss the  Weyl groups of S, T and X dualities, which are subgroups
of the U duality. Section 5 contains an analysis of the multiplet
structure of $p$-brane solutions under the Weyl-group symmetries.  We
shall also obtain a new supersymmetric $p$-brane  solution in $D=3$, which
involves 8 participating 1-form field strengths.  The paper ends with
discussion and conclusions in section 6. 

\section{The U Weyl group in $D\ge 9$}

     We begin this section by discussing the symmetries of $D=9$ 
supergravity.  This theory can be obtained by dimensional reduction of
$D=10$  type IIA supergravity, which in turn can be obtained by
reduction from $D=11$  supergravity.  Alternatively, the same $D=9$
supergravity can be obtained  from the dimensional reduction of type
IIB supergravity in $D=10$.  The two reduction routes result in two
formulations of the $D=9$ theory that are related by an orthogonal
field  redefinition of the dilatonic scalars.  We  shall construct
$D=9$ supergravity here by dimensional reduction of type IIA
supergravity.  However, for convenience, we shall then choose the basis
of dilatonic scalar fields that corresponds to the type IIB reduction
route.  The bosonic sector of the theory contains the vielbein, a
dilaton $\phi$ together with a second dilatonic scalar $\varphi$, one
4-form field strength
$F_4$, two 3-forms $F_3^i$, three 2-forms $F_2^{12}$ and ${\cal F}_2^i$
and one 1-form ${\cal F}_1^{12} = d\chi$.  It follows from
(\ref{dgenlag}), (\ref{dilatonvec}) and (\ref{gfvecb}) that the bosonic
Lagrangian is given by 
\bea {\cal L} &=& eR -\ft12e (\del\phi)^2 -\ft12e (\del\varphi)^2 -
\ft12 e e^{-2\phi} (\del\chi)^2\nonumber\\ && -\ft1{48} e
e^{\ft{2}{\sqrt7}\varphi} (F_4)^2 -
\ft12 e e^{\phi -\ft{1}{\sqrt7}\varphi} (F_3^{(1)})^2 -\ft12 e 
e^{-\phi - \ft{1}{\sqrt7}\varphi} (F_3^{(2)})^2  -\ft14 e
e^{-\ft{4}{\sqrt7}\varphi} (F_2^{(12)})^2 \label{d9lag}\\ && -\ft14 e
e^{-\phi + \ft3{\sqrt7}\varphi} ({\cal F}_2^{(1)})^2 -
\ft14 e e^{\phi +\ft3{\sqrt{7}} \varphi} ({\cal F}_2^{(2)})^2 -\ft12
\td F_4\wedge \td F_4 \wedge A_1^{(12)} -
\td F_3^{(1)} \wedge \td F_3^{(2)} \wedge A_3\ ,\nonumber
\eea where we have defined $\phi=\phi_1$, $\varphi=\phi_2$, and
$\chi={\cal  A}_0^{(12)}$. We are using the notation that field
strengths without tildes include the various Chern-Simons
modifications, whilst field strengths written with tildes do not
include the modifications.   Thus we have: 
\bea F_4&=&\td F_4 - \td F_3^{(1)}\wedge {\cal A}_1^{(1)} - 
\td F_3^{(2)}\wedge {\cal A}_1^{(2)} - \ft12 \td F_2^{(12)} 
\wedge {\cal A}_{1}^{(1)}
\wedge {\cal A}_1^{(2)}\ ,\nonumber\\ F^{(1)}_3 &=& \td F^{(1)}_3 - \td
F_2^{(12)} \wedge {\cal A}_1^{(2)}\ ,
\nonumber\\ F_3^{(2)} &=& \td F_3^{(2)} + F_2^{(12)}\wedge {\cal
A}_1^{(1)} -
\chi (\td F^{(1)}_3 -F_2^{(12)}\wedge {\cal A}_1^{(2)})
\ ,\label{cs9d}\nonumber\\ F_2^{(12)} &=& \td F^{(12)}_2\ ,\qquad {\cal
F}_2^{(1)} = {\cal F}_2^{(1)} +\chi {\cal F}_1^{(2)}\ ,\qquad {\cal
F}_2^{(2)} = \td {\cal F}_2^{(2)}\ , \qquad {\cal F}_1^{(12)} = \td
{\cal F}_1^{(12)}\ .\label{csterms}
\eea In obtaining these expressions from the results in \cite{lp1}, we
have performed the field redefinition ${\cal A}_1^{(1)}\longrightarrow
{\cal A}_1^{(1)} -\chi\,  {\cal A}_1^{(2)}$ in order to arrive at a
formulation where, as we shall see below, $SL(2,\R)$ acts linearly on
the potentials ${\cal A}_1^i$. 

    The Lagrangian (\ref{d9lag}) has a $GL(2,\R) \simeq SL(2,\R) \times
SO(1,1;\R)$ symmetry \cite{cj}.  The $SO(1,1;\R)$ symmetry corresponds to a
constant shift of the dilatonic scalar $\varphi$, together with rescalings
of the gauge potentials: 
\bea &&\varphi\longrightarrow \varphi + 2\sqrt7 c\ ,\nonumber\\ &&A_3
\longrightarrow e^{-2c} A_3\ ,\qquad A_1^{(12)} \longrightarrow e^{4c}
A_1^{(12)}\ ,\nonumber\\ &&A_2^{(i)} \longrightarrow e^{c}\, A_2^{(i)}\
,\qquad {\cal A}_1^{(i)}
\longrightarrow e^{-3c}\, {\cal A}_1^{(i)}\ .
\eea

     In order to understand the action of the $SL(2,\R)$ symmetry,  it
is convenient to re-express the scalar fields $\phi$ and $\chi$, which 
live in the coset $SL(2,\R)/SO(2)$, in terms of three fields $(X,Y,Z)$ 
subject to the $SO(1,2)$-invariant constraint $X^2-Y^2-Z^2=1$:
\bea X&=& \cosh\phi + \ft12 \chi^2 e^{-\phi}\ ,\nonumber\\ Y& = & \sinh
\phi +\ft12 \chi^2 e^{-\phi}\ ,\label{xyz}\\ Z&=&\chi e^{-\phi}\
.\nonumber
\eea The scalar Lagrangian for the  fields $\phi$ and $\chi$ now takes
the simple $SO(1,2)$-invariant form 
\be -\ft12(\del\phi)^2 -\ft12 e^{-2\phi}(\del\chi)^2 =
\ft12 (\del X)^2 -\ft12 (\del Y)^2 - \ft12(\del Z)^2\ .\label{xyzlag}
\ee The group $SO(1,2)$ is isomorphic to $SL(2,\R)$, whose action  on
$(X,Y,Z)$ is
\be
\left( \begin{array}{cc} Z & Y-X \\ Y+X & -Z \end{array}\right)
\longrightarrow \Lambda \left( \begin{array}{cc} Z & Y-X \\ Y+X & -Z 
\end{array}\right) \Lambda^{-1}\ ,
\ee where
\be
\Lambda = \left( \begin{array}{cc} a & b\\c& d\end{array}\right)\ ,
\label{lmatrix}
\ee 
and $ad-bc=1$.  This implies that
\be 
e^{-\phi} \longrightarrow  (a+ b\chi)^2 e^{-\phi} + b^2 e^{\phi}\
,\qquad
\chi e^{-\phi} \longrightarrow (a + b\chi) (c + d\chi) e^{-\phi} + bd\,
e^{\phi}\ ,\label{chiphi}
\ee 
which translates into the fractional linear transformation 
$\lambda\equiv \chi+ i e^\phi \rightarrow (c + d \lambda)/(a+ b
\lambda)$. It is straightforward to see that the entire Lagrangian
(\ref{d9lag}) is invariant under this $SL(2,\R)$ transformation if the
potentials  transform as
\bea A_3 \longrightarrow A_3\ ,&& A_1^{(12)} \longrightarrow
A_1^{(12)}\  ,\nonumber\\ &&\nonumber\\
\left( \begin{array}{c} A_2^{(1)} \\ A_2^{(2)} \end{array}\right)
\longrightarrow \Lambda 
\left( \begin{array}{c} A_2^{(1)} \\ A_2^{(2)} \end{array}\right)\ ,
&&\left( \begin{array}{c} {\cal A}_1^{(1)} \\ {\cal A}_1^{(2)} 
\end{array}\right)
\longrightarrow (\Lambda^T)^{-1}
\left( \begin{array}{c} {\cal A}_1^{(1)} \\ {\cal A}_1^{(2)} 
\end{array}\right)\ .\label{d9ftrans}
\eea This $SL(2,\R)$ symmetry of $D=9$ supergravity can also be
obtained by  dimensionally reducing type IIB supergravity \cite{bho}.

    We may now consider a discrete $S_2$ subgroup of $SL(2,\R)$, 
corresponding to taking $a=d=0$, $b=1$ and $c=-1$.  Under this
permutation group, the non-invariant fields transform as follows:
\bea e^{-\phi} \longrightarrow e^{\phi} + \chi^2 e^{-\phi}\ ,&&
\chi e^{-\phi} \longrightarrow -\chi e^{-\phi} \ ,\nonumber\\ A_2^{(1)}
\longrightarrow A_2^{(2)}\ ,\qquad A_2^{(2)} 
\longrightarrow -A_2^{(1)}\ ,&& {\cal A}_1^{(1)} \longrightarrow {\cal
A}_1^{(2)} \ ,\qquad {\cal A}_1^{(2)} \longrightarrow -{\cal A}_1^{(1)}
\ .\label{d9s2trans}
\eea 
Note that this transformation simply interchanges the indices 1
and 2 of the  pair of 3-form field strengths $\td F_3^{(i)}$, and the
pair of 2-form field  strengths $\td {\cal F}_2^{(i)}$, together with
certain sign changes.  Its  action on the untilded field strengths with
their Chern-Simons  modifications, which appear in the kinetic terms in
(\ref{d9lag}), is more  complicated.  These complications are
associated exclusively with terms involving the  0-form potential
$\chi$.  If we concentrate only on the terms that  are independent of
the bare undifferentiated $\chi$, which can be thought of as the
leading-order terms, we see that the action of this
$S_2$ discrete group is simply to interchange the pairs of field
strengths and at the same time interchange their associated dilaton
prefactors.  This interchange of dilaton vectors is equivalent to $\phi
\longrightarrow -\phi$.  As far as the other field strengths in the
theory are concerned, one can check that the dilaton vectors for the field
strengths $F_4$ and $F_2^{(12)}$ are invariant, since they are independent
of $\phi$.  On the other hand, the $S_2$ transformation reverses the sign of
the dilaton vector for the 1-form field strength ${\cal F}_1^{(12)} =
d\chi$.  The Lagrangian is nevertheless invariant, as we already
demonstrated, since, unlike the higher-degree potentials which transform
linearly, the 0-form potential undergoes the non-linear transformation $\chi
\longrightarrow -\chi e^{-2\phi} + \cdots$, which precisely compensates for
the sign reversal of its dilaton vector. 

     The discrete $S_2$ subgroup (\ref{d9s2trans}) is also seen to coincide
with the subgroup of the $SL(2,\Z)$ U duality group that leaves unchanged
the scalar asymptotic values $(\langle\phi\rangle, \langle\chi\rangle)$ when
these are taken to vanish.  To see this, we observe from (\ref{chiphi}) that
to preserve the vanishing asymptotic values $\langle\phi\rangle =0=
\langle\xi\rangle$, we must have $a^2 + b^2= 1$ and $a c + b d =0$, in
addition to $ad - bc =1$.  Up to a trivial sign change of the matrix 
$\Lambda$ (\ref{lmatrix}), the only solutions are the identity and $a=d=0$, 
$b=-c=1$.  These are precisely the elements of the $S_2$ group that we 
discussed above. This observation gives rise to the alternative
interpretation of this $S_2$ subgroup as the stability duality group of the
vacuum, as presaged in the introduction. 
     
     As we shall see, the pattern that we have described here occurs
also in lower dimensions. In $D$-dimensional maximal supergravity, the
analogous discrete subgroup of the supergravity symmetry group
$G\simeq E_{n(+n)}$ ($n=11-D$) implements certain permutations of 
the dilaton vectors associated with the field strengths. Although the 
9-dimensional example is rather too degenerate to illustrate the point
clearly, we shall see more transparently in lower dimensions, where the
group is larger, that this interchange of the dilaton prefactors corresponds
precisely to the action of the Weyl group $W$ of $E_{n(+n)}(\R)$. The
interpretation of the Weyl group as the stability duality group of the
vacuum will also obtain in the lower-dimensional cases. 

     Now let us consider maximal supergravities in higher dimensions. 
The  symmetry groups of 11-dimensional supergravity and type IIA
supergravity are rather trivial, with no non-trivial Weyl group.  On
the other  hand, type IIB supergravity has an $SL(2,\R)$ symmetry
\cite{sch1}, giving rise to a discrete $S_2$ Weyl-group symmetry.  The
analysis of this Weyl group action is  analogous to that of $D=9$
supergravity.  Since the  self-dual 5-form field strength in type IIB
supergravity is inert under $SL(2,\R)$, we may omit it from the
discussion and consider only the subset of type IIB fields that are
subject to $SL(2,\R)$ transformations. This subset of fields can be
described by a Lagrangian, and in fact since the Lagrangian
(\ref{d9lag}) for the bosonic sector of $D=9$ supergravity is written
in the type IIB basis, we may simply obtain the relevant part of the
type IIB Lagrangian by setting $\varphi$ and the 3-form and 1-form
potentials to zero in (\ref{d9lag}), and by restoring the dependence on
the tenth coordinate. The resulting Lagrangian is therefore invariant
under the $SL(2,\R)$ transformations given by (\ref{chiphi}) and
(\ref{d9ftrans}).  The discrete Weyl-group transformation
(\ref{d9s2trans}) corresponds also to the interchange of the dilaton
vectors of the two 3-form field strengths. 

     It is of interest to note that in addition to the $SL(2,\R)$  symmetry
in type IIB supergravity, there is a further discrete $S_2$  symmetry that
differs from the Weyl-group symmetry discussed above.  Namely, type  IIB
supergravity is invariant under the following discrete transformation: \bea
A_2^{(1)} \longrightarrow - A_2^{(1)}\ ,\qquad A_4 \longrightarrow -A_4\
,\qquad\chi \longrightarrow -\chi \eea with the rest of the fields being
invariant. This is not part of the $SL(2,\R)$ symmetry. Such additional
discrete symmetries lying outside the $G\simeq E_{n(+n)}$ symmetry group
also arise in lower dimensions. 

      So far, we have considered transformations that generate the {\it
symmetries} of the supergravity theories.  In $D=9$ dimensions, there is
another important transformation, namely the $\Z_2$ T-duality of string
theory, which is not a symmetry.  Instead, it relates the type IIA
string compactified on a large circle to the type IIB string
compactified on a small circle. At the level of $D=9$  maximal
supergravity, this is a transformation that maps between the
Lagrangians obtained by dimensional reduction of the type IIA and type
IIB theories. The T-duality transformation corresponds to making the
following field redefinition of the two dilatonic scalars $(\phi,
\varphi)$: 
\be {\left(\begin{array}{c} 
    \phi \\ \varphi
\end{array}\right)}_{IIA}  =\left(\begin{array}{cc}
 \ft34 & -\ft{\sqrt7}{4}\\
                         -\ft{\sqrt7}{4} & -\ft34
                     \end{array}\right) {\left(\begin{array}{c} \phi\\
\varphi \end{array}\right)}_{IIB} \  .\label{tduald9}
\ee This transformation can be derived from the T-duality
transformation in the 
$D=10$ string $\sigma$ model.  The $\sigma$-model transformation that 
corresponds to the transformation (\ref{tduald9}) of the dilatonic
scalar fields  is given by \cite{bus}
\be
\td g^{\sigma}_{zz} = \fft{1}{g^{\sigma}_{zz}}\ ,\qquad \td \phi = 
\phi - \ft12 \log{g^{\sigma}_{zz}}\ ,
\ee 
where $g_{zz}^\sigma$ denotes the diagonal internal component of the
$\sigma$ model metric $g^\sigma_{\sst{MN}}$, upon reduction to $D=9$.
Translating to the Einstein-frame metric $g_{\sst{MN}} = e^{\ft12 \phi}
g^{\sigma}_{\sst{MN}}$ in $D=10$, and performing the dimensional
reduction, which implies $ds^2_{10} = e^{\varphi/(2\sqrt7)} ds_9^2 +
e^{-\sqrt7
\varphi/2} (dz + {\cal A})^2$, we obtain precisely the transformation
(\ref{tduald9}).  One can easily verify that acting with this
transformation on the Lagrangian (\ref{d9lag}) gives rise to the
Lagrangian obtained from dimensional reduction of type IIA
supergravity, whose dilaton vectors are given by (\ref{dilatonvec}) and
(\ref{gfvec}).   Note that if one truncates $D=9$ maximal supergravity
to $N=1$ supergravity, where only the NS-NS fields remain, then this
$T$-duality is actually a symmetry of the theory.

\section{U Weyl group in $3\le D \le 8$}

      We saw in the previous section that there is a discrete $S_2$
symmetry of the $D=9$ maximal supergravity theory, which has the
interpretation of  being the Weyl group of the $SL(2,\R)$ symmetry of
the theory.  In this  section, we shall extend the discussion to the
maximal supergravities in all  dimensions $3\le D \le 8$.  We shall see
that the discrete Weyl-group  symmetry of the theory becomes more
apparent as the symmetry group enlarges  with the descent through the
dimensions.  

\subsection{$D=8$}

    Maximal supergravity in $D=8$ contains one 4-form field strength
$F_4$; three 3-forms $F_3^{i}$; six 2-forms comprising three $F_2^{ij}$
and three 
${\cal F}_2^{ij}$; and four 1-forms comprising one $F_1^{ijk}$ and three
${\cal  F}_1^{ij}$, where the indices run over $1,2,3$.  The dilaton
vectors for these field strengths are given by (\ref{dilatonvec}).  To
discuss the Weyl  group symmetry, we first consider the 4-form and
3-forms.  Since the degree of the 4-form field strength is equal to
$D/2$, it follows that there is an $S_2$ symmetry of the equations of
motion, corresponding to a dualisation of $F_4$, under which its
dilaton vector $\vec a$ reverses its sign.   Under this $S_2$ symmetry,
the dilaton vectors of the 3-forms are inert.  For the 3-forms, there
is however an $S_3$ symmetry, corresponding to permutations of their
dilaton vectors $\vec a_i$, under which the  dilaton vector for the
4-form is invariant.  This independence of the $S_2$ and
$S_3$ symmetries is not immediately manifest.  It can be seen, however,
from the detailed properties of the dilaton vectors. In particular, it
follows from  (\ref{dilatonvec}) and (\ref{gfdot}) that in $D=8$ these
dilaton vectors  satisfy 
\be
\vec a \cdot \vec a_i = 0\ ,\qquad \sum_i \vec a_i = 0\ .\label{vec1}
\ee Thus the vectors $\vec a_i$ form a plane, on which the $S_3$
symmetry acts, whilst the $S_2$ reflects the vectors orthogonal to this
plane.  This 
$S_3\times S_2$ discrete symmetry is isomorphic to the Weyl group of 
$SL(3,\R)\times SL(2,\R)$, which is the symmetry group of $D=8$ maximal 
supergravity \cite{cj}.  In fact, the vectors $\ft1{\sqrt2} \vec a_i$
and
$\pm \ft1{\sqrt2}\vec a$ are precisely the weight vectors of the
fundamental representations of $SL(3,\R)$ and $SL(2,\R)$ respectively.
This can be seen by noting that the weight vectors $\vec h_i$ of  the
fundamental representation of $SL(N,\R)$ satisfy 
\be
\vec h_i \cdot \vec h_j = \delta_{ij} -\fft{1}N\ ,\qquad
\sum_{i} \vec h_i =0\ ,
\ee which is precisely the relation satisfied by $\pm \ft1{\sqrt2} \vec
a$ and $\ft1{\sqrt2} \vec a_i$, for $N=2$ and $N=3$ respectively. These
weight vectors permute among themselves under the actions of the
corresponding $S_3$ and $S_2$ Weyl groups.  

     To complete the discussion of the U Weyl group symmetry in
$D=8$ supergravity  theory, we need to examine its action on the dilaton
vectors of the 2-form  and 1-form field strengths.  These dilaton
vectors can be expressed in terms of $\vec a_i$ and $\vec a$ as follows:
\bea {\rm 2-forms:}&& \vec a_{ij} = -\vec a_k -\vec a\ ,\qquad \vec b_i 
= -\vec a_i + \vec a\ ,\nonumber\\ {\rm 1-forms:}&& \vec a_{123} =
2\vec a\ ,\qquad
\vec b_{ij} = -\vec a_i + \vec a_j\ ,
\eea where in the first line, $k$ takes the value in the set
$\{1,2,3\}$ that is not equal to either $i$ or $j$.   Thus under the
$S_3$ group, the two sets of three 2-form dilaton vectors $a_{ij}$ and
$b_i$ permute among themselves.  Under the $S_2$ group, the two sets
interchange.  For the 1-forms, we  see that the dilaton vector $\vec
a_{123}$ is inert under $S_3$, but has a sign change under $S_2$. The
dilaton vectors $b_{ij}$ permute with sign changes under the $S_3$, and
are inert under the $S_2$.   As we saw in the $D=9$ example in the
previous section, the 0-form potentials, unlike the higher-rank 
potentials, transform non-linearly, in a manner that precisely
compensates for the sign  changes of their dilaton vectors.  Thus, we
see that the six dilaton vectors of the 2-form field strengths form an
irreducible $(3,2)$ representation of the $S_3\times S_2$ Weyl group of
$SL(3,\R)\times SL(2,\R)$.   The dilaton vectors $\vec a_{123}$ and
$\vec b_{ij}$ for the 1-form field strengths, together with their
negatives, transform as the $(1,2)$ and $(6,1)$ representations. 

     Having observed that the dilaton vectors of the various field
strengths form  representations of the Weyl group of $SL(3,\R)\times
SL(2,\R)$, we now argue that the Weyl group is indeed a discrete
symmetry of the theory.   First, note that if we permute the field
strengths appearing in the  Lagrangian (\ref{dgenlag}) at the same time
as their dilaton vectors  are permuted according to the rule given
above, the form of the Lagrangian is  preserved.  Of course, this of
itself does not demonstrate the invariance  of the theory under the
Weyl-group action, since, as we already saw for
$D=9$ in the previous section, the Chern-Simons modifications to the
field  strengths complicate the discussion considerably.  However, as
we also saw  in the previous section,  the Chern-Simons modifications
are sub-leading terms that follow in a prescribed fashion once the
leading-order behaviour  of the unmodified field strengths is
established.  The details of how these  sub-leading terms conspire to
make the theory fully invariant are contained in the standard proof of
the $SL(3,\R)\times SL(2,\R)$ invariance of 
$D=8$ supergravity \cite{cj}.  Thus we may  reduce the discussion of the
Weyl-group invariance to a discussion of the leading-order terms by
invoking these known results, provided we can show that the Weyl group,
as identified in this paper, is contained within the full supergravity
symmetry group
$SL(3,\R)\times SL(2,\R)$.  

     The $S_2$ symmetry reversing the sign of the dilaton vector for the
4-form field strength corresponds to the symmetry of interchanging the
equations of motion and the Bianchi identity of the 4-form field
strength and its dual. This is part of the
$SL(2,\R)$ symmetry.   Under $SL(3,\R)$, the 2-form potentials $A_2^i$
transform as $A_2^i \longrightarrow \Lambda^i{}_j A_2^j$, where
$\Lambda^i{}_j$ is an $SL(3,\R)$ matrix.  It is easy to see that there
is a discrete subgroup of these matrices that permutes the $A_2^{i}$'s,
with certain sign changes as necessary in order to ensure that the
matrices have determinant 1.  This permutation of the 2-form potentials
corresponds precisely to the previously-discussed $S_3$ permutation of
the dilaton vectors of the 3-form field strengths.  The sign changes of
the potentials are unimportant, since the leading-order terms in the
Lagrangian are quadratic in potentials.  This completes the
demonstration of the invariance of
$D=8$ maximal supergravity under the $S_3\times S_2$ U Weyl group.

\subsection{$D=7$}

      There are a total of five 3-form field strengths in $D=7$ maximal
supergravity: four $F_3^{i}$ and  one $F_3 = ^{\ast}F_4$, which is the
dualisation of the 4-form field strength.   The index $i$ runs over
$1,2,3,4$.  The dilaton vectors $\vec a_i$ of the the field strengths
$F_3^{i}$ are given by (\ref{dilatonvec}).  The dilaton vector of
$F_3$, which we denote by $\vec a_5$, is given by $\vec a_5=-\vec a$. 
It  follows from (\ref{dilatonvec}) and (\ref{gfdot}) that these 
vectors satisfy
\be
\ft1{\sqrt{2}} \vec a_{\sst{I}}\,\cdot\,\ft1{\sqrt2}\vec a_{\sst{J}}=
\delta_{\sst{IJ}} - \ft15\ ,\qquad \sum_{\sst I} \vec a_{\sst{I}}=0
\ ,\label{vec2}
\ee where ${\scriptstyle I}=(i,5)$.  These are precisely the defining
properties of the weight vectors of the 5-dimensional fundamental
representation of
$SL(5,\R)$. Thus  these dilaton vectors are permuted and form a
5-component irreducible multiplet under the action of the Weyl group
$S_5$ of
$SL(5,\R)$. 

    Now let us examine how the dilaton vectors of the lower-rank field
strengths transform under this Weyl group.  There are a total of ten
2-form field strengths: six corresponding to the dilaton vectors $\vec
a_{ij}$ and four corresponding to $\vec b_i$.  There are ten 1-form
field strengths: four corresponding to $\vec a_{ijk}$ and six
corresponding to $\vec b_{ij}$. {}From (\ref{dilatonvec}) and
(\ref{vec2}), we see that they can be expressed in terms of $\vec a_i$
and $\vec a_5$ as 
\bea 
{\rm 2-forms:}&&\vec a_{ij} = \vec a_i + \vec a_j + \vec a_5\ ,\qquad
\vec b_i = -\vec a_i -\vec a_5\ ,\nonumber\\ {\rm 1-forms:}&&\vec
a_{ijk} = -\vec a_{\ell} + \vec a_5\ ,\qquad
\vec b_{ij} = -\vec a_i + \vec a_j\ ,\label{f821}
\eea 
where in the second line, $\ell$ takes the value in the set
$\{1,2,3,4\}$  that is different from $i,j$ and $k$.  Owing to the
property that 
$\sum_{\sst{I}} \vec a_{\sst{I}} = 0$, it is easy to see that under the
$S_5$  permutations of $\vec a_{\sst I}$, the dilaton vectors of the
2-form field strengths form a 10-component irreducible multiplet. The
ten  dilaton vectors of the 1-form field strengths, together with their
negatives, form a 20-component multiplet. In fact, up to an overall 
rescaling by
$\sqrt2$, they are precisely the non-zero weights of the  adjoint
representation of $SL(5,\R)$.

      To leading order, the Lagrangian will be invariant under the
$S_5$ Weyl group if we permute the associated field strengths at the
same time as their dilaton vectors.  As in the previous discussions for
$D=9$ and $D=8$, the complete proof of invariance, including the
Chern-Simons modifications, follows by invoking the known $SL(5,\R)$
invariance of $D=7$ supergravity. Noting that the 2-form potentials
$A_2^{\sst I}$ transform under $SL(5,\R)$ as $A_2^{\sst I}{}' =
\Lambda^{\sst I}{}_{\sst J} A_2^{\sst J}$,  we see that there is a
discrete subgroup $S_5$ of matrices that permutes the
$A_2^{i}$'s, with certain sign changes. This precisely corresponds to
the permutation of the associated field strengths in the leading-order
terms of the Lagrangian.

\subsection{$D=6$}

     In $D=6$, we can dualise $F_4$ to a 2-form field strength, and
then the  highest rank of the field strengths is 3.  There are five
3-form field strengths, with corresponding dilaton vectors $\vec a_i$
given in (\ref{dilatonvec}).  Since the 3-form field strengths have
rank $D/2$, there are additional symmetries at the level of the
equations of motion, which interchange the field equations and the
Bianchi identities for the field strengths and their duals.  Such
duality transformations are associated with sign changes of the dilaton
vectors $\vec a_i$.  Thus for the 3-form field strengths considered in
isolation, there is a symmetry consisting of permutations of the $\vec
a_i$ together with an arbitrary number of sign changes.

      Now we examine how this symmetry acts on the dilaton vectors for
the 2-form and 1-form field strengths.  There are a total of sixteen
2-form field strengths, corresponding to dilaton vectors $\vec a_{ij}$,
$\vec b_i$ and $-\vec a$, and there are twenty 1-form field strengths,
corresponding to dilaton vectors $\vec a_{ijk}$ and $\vec b_{ij}$.  The
vectors are given in terms of $\vec a_i$ by
\bea {\rm 2-forms:}&& \vec a_{ij} = \vec a_i + \vec a_j - \ft12
\sum_{k} \vec  a_k\ ,\qquad \vec b_i = -\vec a_i + \ft12 \sum_{k} \vec
a_k\ ,
\qquad -\vec a = -\ft12 \sum_{k} \vec a_k\ ,\nonumber\\ {\rm
1-forms:}&& \vec a_{ijk} =\vec a_i + \vec a_j + \vec a_k -\sum_{\ell} 
\vec a_{\ell}\ ,\qquad \vec b_{ij} = -\vec a_i + \vec a_j\ .\label{vec3}
\eea We may now observe that the full group of permutations and sign
changes of the vectors $\vec a_i$ that we have discussed above does not
map the set of 2-form dilaton vectors into itself.  However, the set
does map into itself if we impose the restriction that the sign changes
must occur in {\it pairs}.  This is precisely the action  of the Weyl
group of $D_5\simeq SO(5,5)$ on the weight vectors of its
10-dimensional fundamental representation.  In fact, the ten vectors
$\pm \ft{1}{\sqrt2} \vec a_i$ are precisely the weight vectors of this
representation, since $\vec a_i\cdot \vec a_j = 2\delta_{ij}$.  The 
dilatons for the 2-form field strengths form a 16-component
irreducible  multiplet of the Weyl group of $SO(5,5)$.  The 20 dilaton
vectors of the 1-form field strengths, together with their negatives,
form a 40-component multiplet of the Weyl group, and they are in fact
$\sqrt2$ times the non-zero weights of the adjoint representation of
$SO(5,5)$.

      Following the same logic as used in the cases 
$D\ge7$, we can show that this U Weyl group action is a symmetry of 
the full $D=6$ supergravity theory.

\subsection{$D=5,4,3$}

      In $D=5$, the 4-form field strength is dualised to a 1-form, and
the  six 3-forms are dualised to 2-forms.  Thus we have a total of 
twenty seven  2-forms, corresponding to 15 dilaton vectors $\vec
a_{ij}$,  6 vectors $\vec b_i$ and 6 vectors $a^{\ast}_i =-\vec a_i$,
where the $\ast$  denotes the dilaton vectors of the dualised 3-forms. 
These 27 dilaton vectors  are given by (\ref{dilatonvec}) and
(\ref{gfdot}), and we find  that in $D=5$, the  vectors $\vec f_i$ and
$\vec g$ can be expressed as
\be
\vec f_i =\sqrt2 \vec e_i +(\ft1{\sqrt6} -\ft1{3\sqrt2})\sum_{j} \vec
e_j
\ ,\qquad \vec g = \sqrt{\ft23} \sum_j \vec e_j\ ,
\ee where $\vec e_i\cdot \vec e_j = \delta_{ij}$.  It turns out that
the 27 dilaton vectors, divided by $\sqrt2$, are precisely the weight
vectors of the 27-dimensional fundamental representation of $E_6$.  To
see this, we note that the simple roots of $E_6$ are given by
\bea
\vec \alpha_1 = \vec e_2 - \vec e_3\ ,\qquad
\vec \alpha_2 = \vec e_3 - \vec e_4\ ,&&
\vec \alpha_3 = \vec e_4 - \vec e_5\ ,\qquad
\vec \alpha_4 = \vec e_4 + \vec e_5\ ,\nonumber\\
\vec \alpha_5 = \ft12(\vec e_1 - \vec e_2 - \vec e_3 -\vec e_4 -\vec
e_5) +\ft{\sqrt3}2 \vec e_6 \ ,&&
\vec \alpha_5 = \ft12(\vec e_1 - \vec e_2 - \vec e_3 -\vec e_4 +\vec
e_5) -\ft{\sqrt3}2 \vec e_6 \ .\label{root5}
\eea The action of the Weyl group of $E_6$ can be generated by the six
Weyl reflections $S_i$ in the hyperplanes orthogonal to the simple roots
$\vec\alpha_i$, whose actions on any vector $\vec \gamma$ are given by
$S_i(\vec\gamma) = \vec \gamma - (\vec \gamma \cdot \vec \alpha_i)
\vec\alpha_i$. The highest-weight  vector $\vec\mu$ of the
27-dimensional fundamental representation is defined  by $\vec \alpha_i
\cdot \vec\mu =\delta_{i6}$. The rest of the 26 weight  vectors can be
obtained by acting with the Weyl group on the highest weight  vector. 
We find that the 27 dilaton vectors for the 2-form field strengths are
$\sqrt2$ times the weight vectors of the 27-dimensional representation
of $E_6$, after an orthogonal transformation of the basis for the unit
vectors $\vec e_i$. 

      There are thirty six 1-form field strengths, corresponding to 20 
dilaton vectors $\vec a_{ijk}$, 15 vectors $\vec b_{ij}$ and one vector
$\vec a^{\ast}=-\vec a$.  These vectors, together with their negatives,
form a  72-component multiplet of the Weyl group of $E_6$.  It is easy
to  verify that they correspond to the non-zero weights of the adjoint
representation of
$E_6$.

     Now let us consider $D=4$ maximal supergravity.  The seven 3-form
field  strengths $F_3^i$ are dualised to 1-form field strengths.  Thus
there are  twenty eight 2-form field strengths, corresponding to $\vec
a_{ij}$ and
$\vec  b_i$, and there are sixty three 1-form field strengths,
corresponding to $\vec a_{ijk}$, $\vec b_{ij}$ and $\vec
a_i^{\ast}=-\vec a_i$.  Since the rank of the  2-form field strengths
is $D/2$, there is a duality symmetry under which their  equations of
motion and Bianchi identities are interchanged. This leads to an
enlarged symmetry group under which the 2-form dilaton vectors and
their negatives are treated on an equal footing.  We find that the 56
vectors $\pm \ft1{\sqrt2} \vec  a_{ij}$, $\pm \ft1{\sqrt2} \vec b_i$
and $\pm \ft1{\sqrt2} \vec a_i$ are the  weight vectors of the 
56-dimensional fundamental representation of $E_7$.  To see this, we follow
the same strategy as used in $D=5$.  First, we  note that the dilaton
vectors are given by (\ref{dilatonvec}) and (\ref{gfdot}), where the vectors
$\vec f_i$ and $\vec g$ can be expressed as 
\be
\vec f_i = \sqrt2 \vec e_i + \ft17(3 - \sqrt2) \sum_j \vec e_j\ ,\qquad
\vec g=\sum_i \vec e_i\ ,
\ee where $\vec e_i\cdot \vec e_j=\delta_{ij}$.  In terms of $e_i$, 
the simple roots of $E_7$ can be written as
\bea
\vec \alpha_1 = \vec e_2 - \vec e_3\ ,&&
\vec \alpha_2 = \vec e_3 - \vec e_4\ ,\qquad
\vec \alpha_3 = \vec e_4 - \vec e_5\ ,\qquad
\vec \alpha_4 = \vec e_5 - \vec e_6\ ,\nonumber\\
\vec \alpha_5 = \vec e_5 + \vec e_6\ ,&&
\vec \alpha_6 = \ft12(\vec e_1 - \vec e_2 - \vec e_3 -\vec e_4 -\vec
e_5  +\vec e_6) -\ft1{\sqrt2} \vec e_7 \ ,\qquad
\vec \alpha_7 = \sqrt2 \vec e_7\ .\label{root4}
\eea The action of the Weyl group of $E_7$ can be generated by the
seven Weyl reflections $S_i$ in the hyperplanes orthogonal to the
simple roots.  The highest weight vector $\vec\mu$ of the
56-dimensional fundamental representation is defined by $\vec \alpha_i
\cdot \vec\mu =\delta_{i7}$.  The rest of the 55 weight vectors are
obtained by acting with the Weyl group on the highest-weight vector. 
We find that the 28 dilaton vectors of the 2-form field strengths,
together with their negatives, are $\sqrt2$ times the weight vectors of
the 56-dimensional representation of $E_7$, after an orthogonal
transformation of the basis for the unit vectors $\vec e_i$.  For the
1-form field strengths, analogously to the higher-dimensional cases,
the 63 dilaton vectors, together with their negatives, form a
126-component multiplet of the Weyl group of $E_7$; they correspond to
the non-zero weights of the adjoint representation of $E_7$. 

      Finally, we consider maximal supergravity in $D=3$, where there
are a  total of 120 1-form field strengths, corresponding to $\vec
a_{ijk}$, $\vec  b_{ij}$, $\vec a_{ij}^{\ast}=-\vec a_{ij}$ and $\vec
b_i^{\ast}= -\vec b_i$.  These vectors are given by  (\ref{dilatonvec})
and (\ref{gfdot}), with $\vec f_i$ and $\vec  g$ given by
\be
\vec f_i = \sqrt2 \vec e_i +\ft{1}{2\sqrt2} \sum_i \vec e_j\ ,\qquad
\vec g = \sqrt2 \sum_i \vec e_i\ ,
\ee where $\vec e_i\cdot \vec e_j=\delta_{ij}$.  To see how the dilaton
vectors transform under the action of the Weyl group of $E_8$, we note
that the simple roots of $E_8$ can be written in terms of $\vec e_i$ as 
\bea
\vec \alpha_1 = \vec e_2 - \vec e_3\ ,&&
\vec \alpha_2 = \vec e_3 - \vec e_4\ ,\qquad
\vec \alpha_3 = \vec e_4 - \vec e_5\ ,\nonumber\\
\vec \alpha_4 = \vec e_5 - \vec e_6\ ,&&
\vec \alpha_5 = \vec e_6 - \vec e_7\ ,\qquad
\vec \alpha_6 = \vec e_7 - \vec e_8\ ,\label{root3}\\
\vec \alpha_7 = \vec e_7 + \vec e_8\ ,&&
\vec \alpha_8 = \ft12(\vec e_1 + \vec e_8 - (\vec e_2 + \cdots +
\vec e_7))\ .\nonumber
\eea It is now very easy to verify that the 120 dilaton vectors,
together with  their negatives, form a 240-component multiplet of the
Weyl group  of $E_8$.  These vectors correspond to the non-zero weights
of the adjoint  representation of $E_8$.

     Thus we see that the dilaton vectors in maximal supergravity
theories  in $3\le D\le 5$ also form finite-order representations of
the Weyl group of the symmetry  group for the corresponding
supergravity theory.  In leading order, these U Weyl group symmetries
correspond to interchanging the potentials (with certain  sign changes)
in parallel with their dilaton vectors.  These  transformations are
discrete subgroups of the full symmetry  groups of the supergravity
theories.  By the same argument as presented in the higher-dimensional
cases, we can see that these Weyl groups are in fact symmetries of the
corresponding supergravity theories.

\section{S, T and X Weyl duality subgroups}

     In the previous two sections, we established that there is a
discrete symmetry in $D$-dimensional supergravity, namely the Weyl
group of the $E_{n(+n)}(\R)$ symmetry group of the supergravity theory, 
where $n=11-D$. $E_{n(+n)}(\Z)$ is the conjectured U duality of the
associated string theory. This U duality contains various subgroups, which
can be identified as S, T and  X dualities. S duality is a conjectured
non-perturbative symmetry of a string theory that that relates the strong
and weak coupling regimes.  In the theories we are considering, namely type
II strings compactified on a torus,  S duality exists only in $D=4$, and has
the effect of interchanging the twelve NS-NS 2-forms with their duals.  It
is an $SL(2,\Z)$ subgroup of the U duality group $E_7(\Z)$. T duality is a
perturbative symmetry of string theory, valid order by order in the string
coupling $g$, although non-perturbative in $\a'$. It is an $SO(n-1,n-1;\Z)$
subgroup of $E_{n(+n)}(\Z)$ that acts on the internal dimensions of the
compactified string, preserving the NS-NS and R-R sectors.  X duality is
another non-perturbative $SL(2,\Z)$ subgroup of $E_{n(+n)}$, which
interchanges the NS-NS and R-R fields. Note that unlike S duality, X duality
exists in all dimensions $D\le 10$ for type IIB strings and $D\le 9$ for
type IIA strings. 

    In this section, we shall study how the Weyl group of the U duality
group
$E_{n(+n)}$ decomposes into Weyl groups of the S, T and X duality
groups.  In  order to do so, it is useful first to identify which of
the field strengths lie in  the NS-NS sector, and which lie in the R-R
sector of the type IIA string.   In $D=10$, the 3-form field strength
is an NS-NS field, and the 4-form and  2-form field strengths are R-R
fields.  In lower dimensions, all fields that  are derived from the
3-form or the metric are NS-NS fields, and all those  derived from the
4-form or 2-form are R-R fields.  Thus, in our notation, we  have
\bea 
{\rm NS-NS:} && \left\{ \begin{array}{ccccc}  F^{(1)}_3 &
F^{(1\a)}_2 & F^{(1\a\beta)}_1 & {\cal F}^{(\a)}_2 & {\cal
F}_1^{(\a\beta)} \\
\vec a_1 & \vec a_{1\a} & \vec a_{1\a\beta} & \vec b_\a & b_{\a\beta}
\end{array} \right. \nonumber\\
{\rm R-R:}&& \left\{ \begin{array}{cccccc} F_4 & F_3^{(\a)} &
F_2^{(\a\beta)} & F_1^{(\a\beta\gamma)} & {\cal F}_2^{(1)} &
{\cal F}_1^{(1\a)} \\
\vec a & \vec a_\a & \vec a_{\a\beta} & \vec a_{\a\beta\gamma} &
\vec b_1 &\vec b_{1\a} \end{array} \right. \label{nsr}
\eea
where we have decomposed the internal index $i$ as $i=(1,\a)$, and
we also indicate the dilaton vector associated with each field strength.

     Let us begin by discussing T duality.  The full T duality group is
the 
$SO(n-1,n-1)$ subgroup of the $E_{n(+n)}$ U duality.  The T Weyl group
is the  Weyl group of $SO(n-1,n-1)$, which is a subgroup of the $U$
Weyl group that  we have discussed in the previous sections.  Since the
dilaton vectors of  the various field strengths form multiplets under
the U Weyl group, we may  examine how they decompose under the T Weyl
subgroup.  We shall show that  dilaton vectors for the NS-NS or the R-R
fields form independent multiplets  under the $T$ Weyl group.  The
details of the multiplet structures depend on  the dimension of the
theory.  However, for the dilaton vectors $\vec  a_{1\a}$ and $\vec
b_\a$ for the $2(10-D)$ NS-NS 2-forms, it can be  discussed in
general.  They form a $2(10-D)$-dimensional representation of 
$SO(10-D, 10-D)$ for $D\le 8$.  To see this, we note that it follows
from  (\ref{dilatonvec}) that these dilaton vectors can be expressed as
\be
\vec a_{1\a} = \sqrt2 \vec e_\a + \ft12(\vec f_1 -\vec g)\ ,\qquad
\vec b_{\a} = -\sqrt2 \vec e_\a + \ft12(\vec f_1 - \vec g)\ ,\label{dnw}
\ee where $\vec e_\a = \ft1{2\sqrt2}(2\vec f_\a + \vec f_1 -\vec g)$,
which satisfy $\vec e_\a\cdot \vec e_\beta = \delta_{\a\beta}$.  The
vectors $\pm 
\vec e_\a$ are the weight vectors of the $2(10-D)$-dimensional
fundamental  representation of $SO(10-D, 10-D)$. The action of the T
Weyl group on the vectors $\vec e_\a$ is to permute any pair, with or
without changing their signs.  Since the vector $(\vec f_1 -\vec g)$ is
invariant under this action, it follows that the vectors $\vec a_{1\a}$
and $\vec b_\a$ form a
$2(10-D)$-component multiplet under the T Weyl group.  This discussion
breaks down in the case of $D=9$, where there is no non-trivial T
duality that interchanges field strengths. Having obtained the weight
vectors of the fundamental representation of the T duality group, on
which the Weyl group action can be simply stated, it is straightforward
to study the multiplet structure of the remaining field strengths,
since their dilaton vectors can be expressed in terms of linear
combinations of $\vec e_\a$ and the T-invariant vector $(\vec f_1 -
\vec g)$.  However, this is not the most convenient way to study the T
Weyl group multiplets for all the field strengths. Since we have
already obtained the U Weyl group multiplet structure in the previous
sections, it is simpler just to read off the T duality structure from
the U duality.  For example, in $D=7$ the U Weyl group
$S_5$ permutes the five 3-form dilaton vectors $\vec a_{\sst I}$, where
$\vec a_5$ is the dilaton vector for the dualised 4-form field
strength. The T Weyl group is the $S_4$ subgroup of $S_5$ generated by
permutations of the 4 dilaton vectors $\vec a_\a$ with $\a = 2,3,4,5$,
{\it i.e.}\ it is the Weyl group of the T duality group $D_3$.  It
follows from (\ref{f821}) that the 6 dilaton vectors for the NS-NS
2-forms and 4 dilaton vectors for the R-R 2-forms comprise 6 and 4
component multiplets under $S_4$ respectively. The 6 dilaton vectors
for NS-NS 1-forms, together with their negatives, form a 12-component
multiplet under $S_4$, and the 4 dilaton vectors for R-R 1-forms
comprise a 4-component multiplet.  Their negatives form another
4-component multiplet.  As discussed in the previous sections, for
$D=9,8,7$ and 6 the U Weyl group is generated by its action on the
dilaton vectors $\vec a$ and $\vec a_i$ for the 4-form and 3-form field
strengths, since these vectors correspond to the weight vectors of the
fundamental representation of the U duality group.  The T Weyl groups,
on the other hand, are generated by the same action, but with the
vector $\vec a_1$ omitted.  (Thus in $D=9$, the T Weyl group is just
the identity.)  For $D=5,4,3$, we showed that the U Weyl groups are
generated by the simple Weyl actions $S_i$, with the roots given by
(\ref{root5}), (\ref{root4}) and (\ref{root3}).  The T Weyl groups are
simply generated by the same simple Weyl actions that permute the
indices $i$ with $i\ne 1$, {\it i.e.}\ they are generated by the simple
Weyl reflections $S_i$ with $i=6$, $1$ and $8$ omitted in $D=5,4$ and 3
respectively. 

      In $D=4$, in addition to the T duality that preserves the NS-NS
and  R-R sectors, there is an $SL(2,\Z)$ S duality that also preserves
the two sectors. The Weyl group of this S duality interchanges the set
of 12 dilaton vectors for the NS-NS 2-form field strengths with their
negatives.  We summarise the T and S Weyl groups and their multiplet
structures for field strengths in the following table: 

\bigskip\bigskip

\centerline{
\begin{tabular}{|c|c|c|c|c|c|c|c|c|c|}\hline Dim. & 
\multicolumn{2}{c|}{3-Forms} & 
\multicolumn{2}{c|}{2-Forms} & \multicolumn{2}{c|}{1-Forms} & T   &  S 
&  U \\ \hline
 &  NS-NS & R-R & NS-NS & R-R & NS-NS & R-R & & & \\ \hline\hline 9 & 
1& 1& 1 + 1& 1& -- & 1 + 1 & $D_1$ & -- & $A_1$ \\ \hline 8 &  (1,1) &
(2,1) & (2,2) & (1,1) + (1,1) &(1,2) + (2,1) & (2,1) + (2,1) & $D_2$ &
-- & $A_2\times A_1$ \\ \hline 7& 1 & 4 & 6 & 4 & 12 & 4 + 4& $D_3$ &
-- & $A_4$ \\ \hline 6& 1 + 1& 8 & 8 & 8 & 24 & 8 + 8 & $D_4$ & -- &
$D_5$ \\ \hline 5& & & 10 + 1 & 16 & 40 & 16 + 16 & $D_5$ & -- & $E_6$
\\ \hline 4& & & (12,2) & (32,1) & (1,2) + (60,1) & (32,2) & $D_6$ &
$A_1$ & $E_7$
\\ \hline 3& & & & & 14 + 14 + 84 & 64 + 64 & $D_7$ & -- & $E_8$ \\
\hline
\end{tabular}}
\bigskip

\centerline{Table 1: T and S Weyl duality multiplets for type IIA
strings}
\bigskip\bigskip

A number of comments on the table are in order.  Firstly, we note that
the various duality groups are given in terms of their Dynkin
classification.  Thus the T duality group in dimension $D$ is 
$D_{n-1}\simeq SO(n-1,n-1)$, where $n=11-D$. The $A_m$ groups appearing
in the S and U duality columns denote the non-compact forms $SL(m+1)$. 
As symmetries of supergravity theories, all the groups considered are
defined over the reals; as conjectured full string symmetries, they are
defined over the integers. In $D=8$, the T duality group is $D_2$,
which is isomorphic to $A_1\times A_1 \simeq SL(2)\times SL(2)$.  The
multiplets under the T Weyl group are specified by their dimensions
under the two $S_2$ Weyl groups of these two
$SL(2)$'s.  The first is the subgroup of the $S_3$ factor in the full U
Weyl group $S_3\times S_2$ that corresponds to permutations just
involving the $\vec a_i$ dilaton vectors with $i=2$ and 3.  The second
$S_2$ is just the $S_2$ factor in $S_3\times S_2$, corresponding to the
interchange of the dilaton vector $\vec a$ and its negative.  In $D=4$,
there is in addition an S duality symmetry.  This arises as the $A_1$
factor in the decomposition of the U duality group down to the T
duality group;
$E_7\longrightarrow D_6\times A_1$.  The orders of the multiplets in
$D=4$ are of the form $(p,q)$, where $p$ is the order under the T Weyl
group of $D_6$, and $q$ is the order under the S Weyl group of
$A_1$.  Note that the occurrence of an extra $A_1$ factor in the
decomposition of the U duality group to the T duality group is unique to
$D=4$, and so toroidal compactifications of the type IIA string exhibit
S duality only for $D=4$.  

     Finally, we remark that there is an X duality in every dimension,
which has the effect of mapping fields between the NS-NS and R-R
sectors.  It is an $SL(2)$ symmetry whose $S_2$ Weyl group is generated
by the Weyl reflection corresponding to the simple root of the U
duality group that is truncated in the passage to the T duality
subgroup. Combined with the
$SL(n-1)$ subgroup of the T duality group that permutes the indices
$\a=\{2,3,\ldots, n\}$ on the dilaton vectors, the X duality has the
effect of interchanging the index $i=1$ with any index $i=\a$, thus
corresponding to an interchange of NS-NS and R-R sectors.  It is worth
remarking T and X dualities do not commute, and that their closure
generates the entire U duality.  Note that we have not included the
dilaton vector $\vec a$ for the 4-form field strength in the table,
since it would appear explicitly only in $D\ge 8$. In fact in $D=9$, it
is a singlet under U duality.  In $D=8$, it, together with its
negative, is a $(1,2)$ under both T and U Weyl dualities.  Under  X
duality, the dilaton vector $\vec a$ and its negative are both 
singlets.

\section{$p$-brane U Weyl group multiplets}     

     In the previous two sections, we established that there is a
discrete  symmetry $W$ in $D$-dimensional supergravity that may be
identified with the Weyl group of the $G\simeq E_{n(+n)}$ symmetry
group for the supergravity theory, where $n=11-D$. Now, we shall study
how $p$-brane solutions form multiplets under the action of the Weyl
group.  First we shall give a short review of
$p$-brane solutions in  maximal supergravity theories, and then we
shall discuss their supersymmetry  properties.  In particular, we shall
show that the supersymmetry properties of a $p$-brane solution are
invariant under the action of the U Weyl group. After this, we shall
discuss the $p$-brane U Weyl  group multiplets.

\subsection{$p$-brane solutions}

       In $D$-dimensional supergravity, one can construct a $p$-brane
solution with $N$ participating $n$-form field strengths, with $N\le
(11-D)$.  The relevant part of the Lagrangian is given by 
\be 
{\cal L} = e R -\ft12 e\, (\del\vec \phi)^2 -\fft1{2n!} e\, \sum_{b
=1}^{N}  e^{\vec c_b\cdot\vec\phi} F_b^2 \ ,\label{genlagn} 
\ee 
where $\vec c_b$ denotes the dilaton vectors associated with the $N$ 
participating field strengths $F_a$. In the single-scalar $p$-brane
solutions, the Lagrangian (\ref{genlagn}) is consistently truncated to the
simple form 
\be 
{\cal L} = e R -\ft12 e\, (\del\phi)^2 -\fft1{2n!} e\, e^{a\phi}
F^2 \ ,\label{genlag}
\ee 
where the scalar field $\phi$ is some linear combination of the dilatonic
scalars $\vec\phi$ of the $D$-dimensional theory, and $F$ is a single
canonically-normalised $n$-index field strength, to which all of the $N$
original field strengths that participate in the solution are proportional.
The constant $a$ appearing in the exponential prefactor can be conveniently
parameterised as 
\be 
a^2 = \Delta - \fft{2d\tilde d} {D-2}\ ,\label{avalue}
\ee 
where $\tilde d\equiv D-d -2$ and $d\tilde d = (n-1)(D-n-1)$.  The
quantity $\Delta$, unlike $a$ itself, is preserved under Kaluza-Klein
dimensional reduction \cite{lpss}.   The value of $a^2$ and the ratios of
the squares  of the field strengths $F_b^2$ are determined by the dot
products of the  dilaton vectors $\vec c_b$.  If the matrix $M_{ab} = \vec 
c_a \cdot \vec  c_b$ is invertible, they are given by \cite{lp1} 
\be 
F_b^2 = a^2 \sum_c (M^{-1})_{bc} F^2 \ ,\qquad  a^2 = \Big( \sum_{b,c}
(M^{-1})_{bc} \Big)^{-1} \ .\label{avaluesol} 
\ee 
If $M_{bc}$ is non-invertible,  there is always a solution with $a=0$, where
the squares of the field strengths $F_b^2$ are equal to each other.  It
turns out that in this singular case, this is the only solution that does
not simply reduce to an already-considered non-singular case with a smaller
number $N$ of participating field strengths \cite{lp1}.  Clearly,  if the
number of participating field strengths exceeds the dimension $(11-D)$ of
the dilaton vectors, then the associated matrix $M_{bc}$ will be
singular, and in fact it turns out that in all such cases, there is no new
solution \cite{lp1}.  Thus in any dimension $D$, it follows that the number
$N$ of participating field strengths must always satisfy $N\le 11-D$. 

     Having reduced the Lagrangian (\ref{dgenlag}) to (\ref{genlag}) by the
above procedure, it is now a simple matter to obtain solutions for the
equations of motion that follow from (\ref{genlag}).  The metric ansatz for
$p$-brane solutions is 
\be ds^2 =e^{2A}\, dx^\mu dx^\nu\eta_{\mu\nu} + e^{2B}\, dy^m dy^m\ ,
\label{metricform}
\ee 
where $A$ and $B$ are functions only of $r=\sqrt{y^my^m}$. The 
coordinates $x^\mu$ lie in the $d$-dimensional world volume of the 
$p$-brane, and $y^m$ lie in the $(D-d)$-dimensional transverse space. 
The  ansatz for the field strength $F$ is either
\be 
F_{m\mu_1\cdots \mu_{n-1}} = \epsilon_{\mu_1\cdots \mu_{n-1}}
(e^C)' 
\fft{y^m}{r}\qquad {\rm or}\qquad F_{m_1\cdots m_n} = \lambda
\epsilon_{m_1\cdots m_np} \fft{y^p}{r^{n+1}}\ .
\label{fansatz}
\ee The first choice yields an elementary $(n-2)$-brane carrying an
electric charge $u=\ft{1}{4\omega_{\sst D-n}}\int_{\del\Sigma}
^{\ast}F$, whilst the second choice yields a solitonic $(D-n-2)$-brane
carrying a magnetic charge $v=\ft{1}{4\omega_n} \int_{\del \Sigma} F$,
where $\del 
\Sigma$ is the boundary $(D-d-1)$-sphere of the transverse space with
volume
$\omega_{\sst D -d -1}$.   The solutions for both the elementary and 
solitonic $p$-branes take the form
\bea ds^2 &=& \Big (1+\fft{k}{r^{\tilde d}}\Big)^{-\ft{4\tilde
d}{\Delta(D-2)}}
\, dx^\mu dx^\nu \eta_{\mu\nu} + \Big(1+ \fft{k}{r^{\tilde d}}\Big)^{
\ft{4d}{\Delta(D-2)}}\, dy^m dy^m\ ,\nonumber\\ e^\phi &=& \Big
(1+\fft{k}{r^{\tilde d}}\Big)^{
\ft{2a}{\epsilon \Delta}}\ , \label{gensol}
\eea where $\epsilon = 1$ and $-1$ for the elementary and the solitonic
solutions respectively, and $k= - \sqrt{\Delta}
\lambda/(2 \tilde d)$.  In the elementary case, the function $C$
satisfies the equation
\be e^C = \fft{2}{\sqrt\Delta} \Big ( 1+ \fft{k}{r^{\tilde
d}}\Big)^{-1}\ .
\ee 
The masses of these solutions are given by $\lambda/(2\sqrt\Delta)$.   Note
that the dual of the field strength in the elementary case is identical to
the field strength of the solitonic case, and {\it vice versa}.  For this
reason, we shall only consider solutions for field strengths with $n\le
D/2$. 

     One way to enumerate all the single-scalar $p$-brane solutions of this
type is to consider, for each dimension $D$ and each degree $n$ for the
field strengths, all possible choices of the associated $N\le 11-D$  dilaton
vectors, and then calculate the values of $a$, and the corresponding ratios
of participating field strengths, using the above equations.   Although this
is easily done for $n=4$ (where there is always only one field strength) and
for $n=3$ (where the number of field strengths is small), for $n=2$ and
$n=1$ the numbers of field strengths grow significantly with decreasing
dimension $D$ and so the enumeration is most conveniently carried out by
computer.  The $p$-brane solutions are characterised by the values of
$\Delta$ and the ratios of the participating field strengths, which are
determined by the set of dot products of the associated dilaton vectors. 
Under the action of the Weyl group, which preserves these dot products,  the
characteristics of the solutions are therefore preserved.  Thus by  acting
with the Weyl group on a given $p$-brane solution, an entire  multiplet of
solutions with the same value $\Delta$ and ratios of field  strengths is
generated. 

\subsection{Supersymmetry and the U Weyl group}

      Another important characteristic of a $p$-brane solution is the 
fraction of 11-dimensional supersymmetry that is preserved by it. We
shall  show explicitly that the U Weyl group also leaves this
characteristic invariant. In order to determine the supersymmetry
properties of the various
$p$-brane solutions, it suffices to study the transformation laws of
$D=11$ supergravity.  In particular, from the commutator of the
conserved supercharges $Q_{\epsilon} = \int_{\del\Sigma} \bar
\epsilon\Gamma^{\sst{ABC}}\psi_{\sst C} d\Sigma_{\sst{AB}}$, we may
read off the $32\times 32$ Bogomol'nyi matrix ${\cal M}$, defined by
$[Q_{\epsilon_1}, Q_{\epsilon_2}]= \epsilon^{\dagger}_1 {\cal M}
\epsilon_2$, whose zero eigenvalues correspond to unbroken generators of
$D=11$ supersymmetry.  The expression for ${\cal M}$ for maximal
supergravity in an arbitrary dimension $D$ then follows by dimensional
reduction of the expression in $D=11$.  A straightforward calculation
shows that it is given by \cite{lp1} 
\bea {\cal M} &=& m\oneone + u\, \Gamma_{012} + u_i\, \Gamma_{01i} +
\ft12 u_{ij}\, \Gamma_{0ij} +  \ft16 u_{ijk} \Gamma_{ijk} + p_i
\Gamma_{0i} + \ft12 p_{ij}\Gamma_{ij} \nonumber\\ &&+
v\,\Gamma_{\hat1\hat2\hat3\hat4\hat5} + v_i \,
\Gamma_{\hat1\hat2\hat3\hat4i}+\ft12 v_{ij}\, \Gamma_{\hat1\hat2\hat3ij}
+ \ft16 v_{ijk}\, \Gamma_{\hat1\hat2ijk} + q_i\,
\Gamma_{\hat1\hat2\hat3 i} +\ft12 q_{ij}\, \Gamma_{\hat1\hat2ij}\ .
\label{genbog}
\eea 
The indices $0, 1,\ldots$ run over the dimension of the $p$-brane
worldvolume, $\hat1,\hat2,\ldots$ run over the transverse space of the $y^m$
coordinates, and $i,j,\ldots$ run over the dimensions compactified in the
Kaluza-Klein reduction from 11 to $D$ dimensions. The quantities $u$, $u_i$,
$u_{ij}$, $u_{ijk}$, $p_i$ and $p_{ij}$ are the electric charges, and $v$,
$v_i$, $v_{ij}$, $v_{ijk}$, $q_i$ and $q_{ij}$ are the magnetic charges,
associated with $F_4$, $F_3^i$, $F_2^{ij}$, $F_1^{ijk}$, ${\cal F}_2^i$ and
${\cal F}_1^{ij}$ respectively. These magnetic and electric charges are 
given by integrals of the corresponding field strengths and their duals 
respectively, together with Chern-Simons corrections, over the boundary of 
the transverse space of the solution \cite{ds,page}.  The quantity $m$ is
the mass per unit $p$-volume for the solution, given by $\ft12 (A'-B')e^{-B}
r^{\td d +1}$ in the limit $r\rightarrow \infty$, where $\td d\equiv D-d-2$.

     Once the mass per unit $p$-volume and the Page charges have been
determined for a given $p$-brane solution, it becomes a straightforward
algebraic exercise to substitute the results for the Page charges into
the Bogomol'nyi matrix (\ref{genbog}) and to calculate the eigenvalues.
The fraction of $D=11$ supersymmetry that is preserved by a solution is
then equal to $k/32$, where $k$ is the number of zero eigenvalues of the
$32\times32$ \bog matrix.  We shall show that the Bogomol'nyi matrix is 
invariant under the action of the U Weyl group, and that $p$-brane 
solutions with the same set of eigenvalues form representations of the U
Weyl group.

     In our discussion of the U Weyl groups in all dimensions $D\le 9$,
we saw  that they always have a subgroup $S_{11-\sst D}$ consisting
simply of the permutations of the internal compactified dimensions,
{\it i.e.}\
$i\leftrightarrow j$ for any $i$ and $j$.  This symmetry can be easily
seen in the \bog matrix by considering the transformation ${\cal M}
\longrightarrow (L_{ij})^{-1} {\cal M} L_{ij}$, where 
\be L_{ij} = e^{\ft{\pi}4 \Gamma_{ij}} = \ft1{\sqrt2}(\oneone +
\Gamma_{ij})\ .
\ee This transformation interchanges the internal index values $i$ and
$j$ on  the $\Gamma$ matrices appearing in ${\cal M}$ (with sign
changes in some  cases).  Thus the \bog matrix is invariant provided
that the indices of the Page charges undergo appropriate conjugate
transformations.  In the case of
$D=9$,  where the non-trivial symmetry group is $SL(2,\R)$, this $S_2$
permutation symmetry is in fact the same as the entire Weyl-group
symmetry of the theory.  Thus the \bog matrix, and hence the
supersymmetry properties of
$p$-brane solutions, are invariant under the action of the U Weyl
group.  It is easy to check that the sign changes of the Page charges
are precisely consistent with the sign changes of the potentials given
in (\ref{d9s2trans}). 

In lower dimensions, $D\le8$, the U Weyl group is larger than the
$S_{11-\sst D}$ permutation group, due to additional discrete
symmetries that involve  the interchange of certain field strengths and
the Hodge duals of other field  strengths.  The rank of the
supergravity  symmetry group is $(11-D)$, which implies that its Weyl
group should be  generated by $(11-D)$ group elements.  We have already
obtained $(10-D)$  independent generators, namely $L_{i,i+1}$ with
$1\le i \le (10-D)$, from  which the entire set of $L_{ij}$ group
elements can be constructed.  The  remaining independent generator can
be taken to be $L_{123}$, where 
$L_{ijk}$ is defined as\footnote{In each dimension $D$, the matrices 
$\Gamma_{ij}$ and $\Gamma_{ijk}$ form some (or in $D\ge 6$ all) of the 
generators of the automorphism group of the associated Clifford
algebra.  Thus in $D=9$, $\Gamma_{12}$ alone closes on $SO(2)$, whilst
in $D=8$, 7 and  6, $\Gamma_{ij}$ and $\Gamma_{ijk}$ close on
$SO(3)\times SO(2)$, $SO(5)$ and 
$SO(5)\times SO(5)$ respectively. In $D=5$ and $D=4$, we must include 
$\Gamma_{ijk\ell mn}$ as well, thereby achieving closure on $USp(8)$
and 
$SU(8)$ respectively.  In $D=3$ we must add the further generators
$\Gamma_{ijk\ell mnp}$ too, thereby achieving closure on $SO(16)$.}
\be L_{ijk} = e^{\ft{\pi}4 \Gamma_{ijk}} = \ft1{\sqrt2} (\oneone +
\Gamma_{ijk})\ .
\ee It is straightforward to establish that the \bog matrix is also
invariant under  the transformation ${\cal M}\longrightarrow
(L_{ijk})^{-1} {\cal M}  L_{ijk}$, provided again that the necessary
conjugate transformations of the  Page charges are performed.  The
complete finite group of transformations generated by $L_{i,i+1}$ and
$L_{123}$ is in fact isomorphic to the U Weyl group discussed
previously, now implemented on the $\Gamma$ matrices appearing in the
\bog matrix.  This can be illustrated simply by concentrating, in each
dimension $D\le 8$, on the group action on the set of Page charges of
the highest-degree field strengths in that dimension. 

     In $D=8$, the transformations generated by $L_{ij}$ form the $S_3$
subgroup of the $S_3\times S_2$ Weyl group.  On the other hand,
$L_{123}$ interchanges $\Gamma_{012}$ and
$\Gamma_{\hat1\hat2\hat3\hat4\hat5}$ in the
\bog matrix, and therefore corresponds to the $S_2$ symmetry that
interchanges the electric and magnetic charges of the 4-form field
strength. It is easy to verify that the lower-degree Page charges also
transform in the proper way, as dictated by the Weyl group.  In $D=7$,
$L_{ijk}$ can also be expressed as $L_{i8}=e^{\ft{\pi}4 \Gamma_{i8}}$
where $\Gamma_8$ is the product $\prod_{\mu} \Gamma_\mu\prod_m\Gamma_m$
of the $D=11$ $\Gamma$ matrices with 7-dimensional spacetime indices.
The action of $L_{i8}$ on the $\Gamma$ matrices corresponds to
interchanging the Page charge of the dual of the  4-form field strength
$F_4$ with the Page charge of the 3-form field strength $F^i_3$. Thus
the 4-form and 3-form sector of the \bog matrix is invariant under the
extended $S_5$ Weyl group of $SL(5,\R)$.   A more detailed analysis
shows that the lower-degree sectors of the \bog matrix are also
invariant under the $S_5$ Weyl group. In $D=6$, the group elements
$L_{ijk}$ can be expressed as $L_{ij7}$, where $\Gamma_7$ is the
product of the $D=11$ $\Gamma$ matrices with $6$-dimensional spacetime
indices. The action of $L_{ij7}$ is to interchange the electric and
magnetic charges of the field strengths $F_3^i$ and $F_3^j$.  Together
with the action of
$L_{ij}$, this has the effect of interchanging the dilaton vectors
$\vec a_i$ and $\vec a_j$ with or without sign change.  This is
precisely the action of the Weyl group of $D_5\simeq SO(5,5)$, as
discussed earlier.

     The analysis of Weyl group invariance of the \bog matrix is more
complicated in the lower dimensions $D\le 5$, since now the symmetry
groups  are enlarged to the exceptional groups $E_6$, $E_7$ and
$E_8$.    In $D=5$ and $D=4$, we can choose one of the $\Gamma$
matrices appearing in 
${\cal M}$ that corresponds to any one of the Page charges for 2-forms,
and fill out the entire set of 27 or 56 such $\Gamma$ matrices
respectively, by  acting with the generators $L_{i,i+1}$ and
$L_{123}$.  Since this translates  into a conjugate action on the Page
charges, it follows that if we start from one 2-form Page charge, we
can fill out the complete 27 or 56-component multiplets.  This is
consistent with what we saw in the previous section, where we showed
that the dilaton vectors for the 2-form field strengths in $D=5$ and
$D=4$ form the 27 and 56 irreducible representations of the relevant
Weyl groups.   In $D=3$, only 1-form field strengths need be
considered, since 2-forms are dual to 1-forms. In total there are 120
1-forms.  So far, we have avoided discussing explicitly the Weyl group
action on the 1-form sector of the \bog matrices, since the 0-form
potentials undergo a more complicated non-linear transformation.  In
the previous section we saw that the full set of $\ell$ dilaton vectors for
the 1-forms, together with their negatives, form a $2\ell$-component
multiplet of the Weyl group.  The reason why these dilaton vectors can
change sign is quite different from the sign changes that we have seen for
the dilaton vectors of the higher-degree forms in the special dimensions
where their duals are of the same degree as the forms themselves.  In those
cases, changing the sign of a dilaton vector was associated with an
interchange of the electric and magnetic charges of the associated field
strength. Here, on the other hand, the sign change of the dilaton vector
$\vec c$ for a 1-form field strength can occur in any dimension, and is a
consequence of the non-linear nature of the transformation for the 0-form
potential, $A_0\longrightarrow A_0'\sim e^{\vec c\cdot \vec\phi}
A_0+\cdots$, which compensates for the sign change of the dilaton prefactor
in the Lagrangian. The Page charge remains invariant under this
transformation, and so the set of $\ell$ 1-form Page charges form an
$\ell$-component multiplet of the Weyl group, even though their dilaton
vectors form a $(2\ell)$-component multiplet.  This is consistent with the
fact that in all dimensions, the action of the generators $L_{i,i+1}$ and
$L_{123}$ on any one of the $\Gamma$ matrices for the 1-forms fills out the
full set of $\Gamma$ matrices for the 1-form field strengths. 

\subsection{$p$-brane U Weyl group multiplets}     

     The procedure for obtaining a $p$-brane multiplet of the U Weyl
group is now very simple.  Consider a $p$-brane solution involving $N$ field
strengths $F_a$ whose dilaton vectors are $\vec c_a$ with $c = 1, \cdots,
N$. Then acting with the U Weyl group on the set of vectors $\{\vec c_1,
\vec c_2, \cdots,\vec c_{\sst N}\}$ generates a full representation of the
Weyl group, where each set of dilaton vectors has the same dot products as
the  original set.  It is then straightforward to identify the associated
sets of participating field strengths.  The $p$-brane solutions constructed
with these sets of field strengths form a multiplet under the U Weyl group.

    The simplest multiplet structures occur for $p$-brane solutions
involving only one field strength.  These solutions all have $\Delta=4$, and
preserve $\ft12$ of the 11-dimensional supersymmetry. In $D=9$, the
$p$-brane solutions using $F_4$ or $F_2^{(12)}$ are singlets, while those
using $F_3^i$ or ${\cal F}_2^{i}$ are doublets.  The $p$-brane solution
using ${\cal F}_1^{12}$ and its Weyl group partner (where the sign of its
dilaton vector is reversed, and the 0-form potential undergoes a non-linear
transformation) also form a doublet.  The analogous multiplet structures can
also be obtained in lower dimensions. For example, in $D=4$, acting with the
Weyl group of $E_7$ on any purely electric or purely magnetic 0-brane
solution fills out a 56-component multiplet.  In fact these are the 56
purely electric or purely magnetic black hole solutions in $D=4$, whose 
significance was addressed in \cite{ht}.  Analogously, acting on any string 
solution using a 1-form field strength fills out a 126-component  multiplet
of the Weyl group of $E_7$. This representation corresponds  to the Weyl
group action on the 126 non-zero weights of the adjoint  representation of
$E_7$.  (Note that solutions related simply by sign changes of the electric or 
magnetic charges are regarded as being equivalent.)

   The U Weyl group multiplet structures of $p$-brane solutions
involving more than one field strength are more complicated.  For
solutions with $N=2$ field strengths, there are two possible values of
$\Delta$, namely $\Delta = 2$ and $\Delta=3$.  The solutions with
$\Delta = 2$ preserve $\ft14$ of the supersymmetry, whilst those with
$\Delta=3$ break all the supersymmetry. The solutions with different
$\Delta$ values cannot lie in the same multiplet since the dot products
of the participating dilaton vectors are different. This is consistent with
our discussion in the previous section, where we showed that the Weyl group
preserves the \bog matrix, and hence it preserves the supersymmetry
properties of the solutions.  In $D=9$, there is just one $p$-brane solution
involving the two 3-form field strengths.  This solution has $\Delta=3$. 
There is also a singlet $\Delta=3$ solution using the two 2-forms ${\cal
F}_2^{i}$, and a doublet of $\Delta=2$ solutions using $F_2^{(12)}$ together
with either of the ${\cal F}_2^{i}$.  Another example is $D=4$. We find in
this case that there are a total of 1512 solutions, half of which have
$\Delta=2$, and the other half have $\Delta=3$.  

       With increasing numbers of participating field strengths, the
possible values of $\Delta$ proliferate.  The discussion of the
multiplet structures for these solutions becomes tedious, albeit
straightforward.   However, most of these solutions are non-supersymmetric,
and we shall not discuss them here in  detail. The supersymmetric solutions
have been classified in \cite{lp1}.   They have $\Delta=\ft{4}{N}$ for
solutions with $N$ participating field strengths, with the corresponding
Page charges having equal magnitudes. There is only one 4-form field
strength, which exists for $D\ge 8$ (in the sense that in dimensions lower
than 8, it will be dualised to a lower-degree form).  The corresponding
$p$-brane solutions are singlets for $D\ge 8$, and form a doublet in $D=8$.
The 3-form field strengths exist for $D\ge 6$. The supersymmetric solutions
all involve one field strength only, and thus have $\Delta=4$. They form
irreducible multiplets of order $2,3,5$ and 10 under the relevant Weyl
groups in $D=9,8,7$ and 6 respectively.  The $p$-brane solutions involving
more than one field strength are all non-supersymmetric, and the values of
$\Delta$ are given by $\Delta = 2 + \ft{2}N$, with $N$ from 2 up to 5.  The
multiplet structures of the 3-form $p$-brane solutions are presented in the
following table: 

\bigskip\bigskip
\centerline{
\begin{tabular}{|c|c|c|c|c|c|}\hline Dim. 
 &\phantom{for} $\Delta=4$\phantom{for} &\phantom{for} 
$\Delta=3$\phantom{for}  &\phantom{for}$\Delta=\ft83$\phantom{for} &
$\Delta=\ft52$ &
$\Delta=\ft{12}5$ \\ \hline\hline
$D=9$    & 2  & 1    &  & & \\ \hline
$D=8$    & 3   &  3   & 1 & & \\  \hline
$D=7$    & 5   & 10   & 10  &1& \\  \hline
$D=6$    & 10   & 40  & 80 & 80 & 32 \\  \hline
\end{tabular}}
\bigskip
\centerline{Table 2: U Weyl multiplets for 3-form solutions} 
\bigskip

\noindent{The 2-form field strengths exist for $D\ge 4$.  The
supersymmetric solutions can involve up to 4 participating field
strengths.  The multiplet structure for these is given in the table
below:} 

\bigskip\bigskip
\centerline{
\begin{tabular}{|c|c|c|c|c|}\hline Dim. 
 &\phantom{for} $\Delta=4$\phantom{for} &\phantom{for}
$\Delta=2$\phantom{for}  &\phantom{for}$\Delta=\ft43$\phantom{for} &
$\Delta=1$\\ \hline\hline
$D=9$    & 1 + 2   & 2    &  &  \\ \hline
$D=8$    & 6   &  6   &  &  \\  \hline
$D=7$    & 10   & 15    &  & \\  \hline
$D=6$    & 16   &  40  &  &  \\  \hline
$D=5$    & 27   &  135   &  45 & \\  \hline
$D=4$    & 56   & 756  & 2520 & 630\\
\hline
\end{tabular}}
\bigskip
\centerline{Table 3: U Weyl multiplets for supersymmetric 2-form
solutions} 
\bigskip

\noindent{The 1-form field strengths exist for all $3\le D \le 9$.  The
supersymmetric solutions can involve up to 8 participating field
strengths.  The multiplet structure for them is given in the table
below:}

\bigskip\bigskip
\centerline{
\begin{tabular}{|c|c|c|c|c|c|c|c|c|c|}\hline Dim. 
 &$\Delta=4$&
$\Delta=2$ & $\Delta=\ft43$ & $\Delta=1'$ & $\Delta=1$ &
$\Delta = \ft45$ & $\Delta=\ft23$ & $\Delta = \ft47$ & 
$\Delta=\ft12$ \\ \hline\hline
$D=9$    & 2   &    &  &  & & & & &\\ \hline
$D=8$    & 8   & 12    &  &  &&&&&\\  \hline
$D=7$    & 20 &  60  &  & &&&&& \\  \hline
$D=6$    & 40  & 280   & 480 & 240 &&&&& \\  \hline
$D=5$    &  72  &  1080   &  4320 & 2160 &&&&&  \\  \hline
$D=4$    & 126   & 3780  & $\begin{array}{c} 30240\\ +2520 \end{array}$ 
& 15120 & 60480 & 90720 & 60480 & 17280 &\\ \hline
$D=3$ & 240 & 15120 & 302400 & * & * & * & * &* &* \\ \hline
\end{tabular}}
\bigskip
\centerline{Table 4: U Weyl multiplets for supersymmetric 1-form
solutions} 
\bigskip

\noindent Note that in $D=4$, the 1-form solutions with $\Delta=\ft43$
form two distinct irreducible multiplets under the U Weyl group, even
though they have the same amount of preserved supersymmetry.  Such a
phenomenon occurs also for the 2-form solutions with $\Delta=4$ in
$D=9$.  In all other cases, the U Weyl group multiplets for
supersymmetric solutions occur in single irreducible multiplets.  Note
also that there are two different 1-form solutions  with $\Delta=1$ in
$D=4$, which we denote by $\Delta=1'$ and $\Delta=1$
\cite{lp1}. The former preserves $\ft18$ of the supersymmetry and the
latter preserves $\ft1{16}$. They form different  irreducible
multiplets, as indicated in table 4.  

     The multiplicities for 1-form solutions in $D=3$ are large, and we 
have obtained results only for $N\le 3$ field strengths; the entries 
containing a $*$ in table 4 correspond to solutions whose multiplicities
have not yet been determined.  Since there are 120 1-form field
strengths, it follows that $\Delta=4$ solutions form a 240-component
multiplet of the Weyl group of $E_8$.  There can be up to $N=8$
participating field strengths, and $\Delta = \ft{4}{N}$, with the new
value $\Delta = \ft12$ occurring when $N=8$. The \bog matrices, and hence
the amount of preserved supersymmetry, for the solutions with $N\le 7$ are
the same as those in $D=4$, as discussed in \cite{lp1}. For the
supersymmetric solutions with 8 field strengths, whose dilaton vectors
satisfy $\vec c_a\cdot \vec c_b =4\delta_{ab}$, we find that the solution
preserves $\ft1{16}$ of the supersymmetry. 

    The above discussion of the Weyl-group multiplet structure for
$p$-brane solutions can be applied not only to the single-scalar
solutions with purely electric or purely magnetic charges, but also to
multi-scalar solutions, and to dyonic solutions in even dimensions. 
Each supersymmetric solution involving $N$ field strengths can be
generalised to an N-scalar solution, in  which the $N$ Page charges
become independent parameters \cite{lp2}.  These multi-scalar solutions
form a bigger multiplet than the single-scalar  ones, since the Page
charges are now independent and hence the participating  field
strengths are distinguishable.  For example, for solutions with 
$N$ 2-form field strengths, the dimensions of the Weyl group multiplets
are increased by a factor of $N!$.  In $D=3$, a new supersymmetric
solution arises with 8 field strengths, which has not been discussed
previously.  The  metric of this multi-scalar solution is given by the
general results in 
\cite{lp2}.  We find that the eigenvalues of the \bog matrix are given
by
\bea
\mu &=& 2\{0, \lambda_{1245}, \lambda_{1346}, \lambda_{2356}, 
\lambda_{1237}, \lambda_{3457}, \lambda_{2467}, 
\lambda_{1567}, \lambda_{2348}, \lambda_{1358},\nonumber\\ &&\ \
\lambda_{1268}, \lambda_{4568}, \lambda_{1478}, \lambda_{2578}, 
\lambda_{3678}, \lambda_{12345678}\}\ ,\label{d3eig1}
\eea 
where $\lambda_{a\cdots b} = \lambda_a + \cdots + \lambda_b$, and
$\ft14\lambda_a$ are the eight independent Page charges for the field
strengths, for example ${\cal F}_1^{(12)}$, ${\cal F}_1^{(34)}$, ${\cal
F}_1^{(56)}$, $F_1^{(127)}$, $F_1^{(347)}$, $F_1^{(567)}$,
$^{\ast}F_2^{(78)}$ and $^{\ast}{\cal F}_2^{(8)}$.  Each eigenvalue in 
(\ref{d3eig1}) has degeneracy 2. Note that the last of the eigenvalues 
(\ref{d3eig1}) is twice the mass of the $p$-brane solution. When all the
$\lambda_a$ are the same, the solution reduces to the single-scalar solution
with $\Delta=\ft12$, and the eigenvalues become $\mu = m\{ 0_2, 1_{28},
2_2\}$.  Thus it preserves $\ft1{16}$ of the supersymmetry, as for the case
with generic values of the Page charges. There are further supersymmetry
enhancements for special values of the charges $\lambda_a/4$, as can be seen
from (\ref{d3eig1}).  In some cases, the \bog matrix has indefinite
signature while in other cases all the non-zero eigenvalues are positive. As
in some of the previous $p$-brane solutions with $N\ge 4$ field strengths,
the supersymmetry of this $N=8$ solution is also sensitive to the choices of
signs for the Page charges $\lambda_a/4$, with half of the sign choices
giving the supersymmetric solution we described above.  The other half of
the choices give rise to solutions that are non-supersymmetric for generic
values of the Page charges.  The eigenvalues of the \bog matrix are given by
\bea
\mu &=& 2\{\lambda_{8}, \lambda_{234}, \lambda_{135}, \lambda_{126}, 
\lambda_{456}, \lambda_{147}, \lambda_{257}, \lambda_{367},
\lambda_{12458}, 
\lambda_{13468}, \nonumber\\ &&\ \ \lambda_{23568}, \lambda_{12378},
\lambda_{34578}, \lambda_{24678}, 
\lambda_{15678}, \lambda_{1234567}\}\ .
\eea 
Note that for generic values of the charges, none of the eigenvalues
is proportional to the mass.

    Now let us consider the dyonic solutions, which exist in $D=8,6$
and 4. There are two kinds of dyonic solutions.  In the first type,
although there are both electric and magnetic charges, each
participating field strength  carries either electric charge or
magnetic charge, but not both.  These types of solutions are already
present in the multiplets discussed above, in $D=6$ and 4 for the cases
with more than one field strength.  This is simply because the dilaton
vectors for 3-forms in $D=6$ (and 2-forms in $D=4$), together with their
negatives (corresponding to dualisation of the fields), form an irreducible
multiplet under the Weyl group.  Thus if we start with a solution with
purely electric or purely magnetic charges, the action of the Weyl group
will map it to a set of solutions including some with both electric and
magnetic charges.  In the dyonic solutions of the second type, each
participating field strength can be both electric and magnetic.  In $D=8$,
such a dyonic membrane solution, which involves a non-vanishing axion also,
was constructed in \cite{ilpt}.\footnote{Commonly, when $p$-brane solutions
are constructed in the literature, the equations of motion are simplified by
considering cases where the ${\cal L}_{\sst{FFA}}$ terms and the
Chern-Simons modifications to the field strengths make no contribution.  In
the multiplets of solutions that we have discussed, it sometimes happens
that some members of a multiplet involve field configurations for which
these contributions do not vanish. In these cases, certain 0-form potentials
become non-vanishing.  Note, however, that the associated 1-form field
strengths do not carry any electric or magnetic charge.}  The action of the
$S_2$ factor of the $S_3\times S_2$ Weyl group is to interchange the
electric and magnetic charges of the 4-form field strength, and hence they
form a doublet.  In $D=6$, dyonic solutions involving only one 3-form field
strength, and with no non-vanishing axions, form a 10-component multiplet
under the Weyl group.  Dyonic solutions involving more than one 3-form field
strength have not yet been constructed. These solutions in general would be 
non-supersymmetric since the corresponding solutions with purely electric or
purely magnetic charges are non-supersymmetric. In $D=4$, examples of dyonic
solutions of the second type with $N=2$ and $N=4$ field strengths have been
constructed in \cite{lp1}.  These solutions involve no non-vanishing axions.
 The U Weyl group action on these solutions generates multiplets of orders
756 and 630, just as for the purely electric or purely magnetic solutions. 
Some members of the multiplets have non-vanishing axions.  Dyonic solutions
of the second type with $N=1$ and 3 field strengths have not yet been
constructed, since all those solutions seem to involve non-vanishing axions.
The analysis of their multiplet structure would, however, be
straightforward. 

\section{Conclusions}

       In this paper, we have studied the Weyl groups of the U duality 
groups for the type IIA string compactified to $D$ dimensions on a
torus.   The U Weyl group describes a discrete symmetry of the theory
that corresponds  to certain permutations of the field strengths, and
in some cases their  duals.  It is analogous to the discrete $\Z_2$
subgroup of the $U(1)$ duality symmetry in electromagnetism, which
interchanges the roles of $B$ and $E$,  and which captures the essence
of the electric-magnetic duality. Although the U duality group, like
its discrete Weyl group, preserves the  isotropicity of a $p$-brane
solution, the full U duality group will transform a solution with $N$
participating field strengths into a solution with more than $N$ field
strengths involved, and with a more complicated axion configuration. 
This complexity simply reflects the fact that an  intermediate rotation
has been performed, in which the resulting  configuration is no longer
aligned cleanly along the axes in the space of  field strengths.

     We have also discussed the Weyl groups of various subgroups of the
U duality  group, namely the S, T and X duality subgroups.  The Weyl
groups of  the S, T and X duality groups single out the subsets of these
transformations that keep the solutions aligned along the axes in the
space of field strength tensors.

     We have seen that the U Weyl groups may also be identified as the
stability duality groups of the vacuum. This is most easily seen by
restricting attention to the standard vacuum, where the scalar fields
asymptotically tend to zero. For this standard vacuum, the full U
duality group $G(\Z)\simeq E_{n(+n)}(\Z)$ is simply represented by
integer-valued matrices. This identification is made without implying a
commitment to any particular interpretation of duality symmetries as
either ordinary global symmetries, or as effectively local discrete
symmetries, to be divided out in constructing the string spectrum. If
one chooses the former view, dualities may be seen as generators of
multiplets of distinct, {\it i.e.}\ non-identified,
$p$-brane solutions in the fashion of Ref.\ \cite{ht}. In this view,
the action of the $G(\Z)$ transformations on $p$-brane solutions
generally needs to be followed by an analytic continuation of the
moduli ({\it i.e.}\ the asymptotic values of the scalar fields) back to
their standard values, so as to remain within the original vacuum
sector. This analytic continuation of the moduli does not preserve the
tensions ({\it i.e.}\ the masses/unit $p$-volume) of the solutions,
giving rise to a $G(\Z)$ multiplet of solutions at different tensions.
However, since the Weyl subgroup of $G$ preserves the standard vacuum,
no such analytic continuation is necessary. Thus, the Weyl group may
also be seen as the symmetry group of $p$-brane solutions at fixed
tension, within a given vacuum sector.

     We have seen that the U Weyl group preserves the number of field
strengths that participate in a given solution.  A remark should be made
about the role of the axions at this point.  Axions play two different roles
in the various $p$-brane solutions.   Firstly, an axion can arise as a
0-form potential for a 1-form field strength  carrying an electric or
magnetic charge, giving rise to an instanton or $(D-3)$-brane solution.
Secondly, an axion can instead arise as a field that is required to be
non-zero in a solution where some other field strengths carry electric
and/or magnetic charges.  In this latter case, the field strength of the
axion itself does not carry any electric or magnetic charge. In a solution
with only one higher-degree ($n\ge 2$) field strength, all the axions
vanish, and they remain zero under any U Weyl group transformation.  If the
solution involves only one 1-form field strength, the U Weyl group
transforms it to a solution with another 1-form field strength, with the
rest of the axions vanishing.  The axions play their second  role in
solutions with more than one field strength, after acting with the U Weyl
group transformation; however, as we stated above, the corresponding 1-form
field strengths do not carry electric or magnetic charges. Thus to be
precise, the U Weyl group preserves the number of non-vanishing charges in a
$p$-brane solution, while the full U duality group does not. 

     The results of this paper may be combined with the various
processes of dimensional  reduction and oxidation to establish
relations between 'brane solutions of different dimensionality $p$.
There are two basic types of dimensional reduction relevant to the
classification of $p$-branes. One is diagonal dimensional reduction
\cite{dhis,lpss}, proceeding by identification of points along the
orbits of isometries of the solutions tangential to the worldvolume.
The second is ``straight'' dimensional reduction \cite{ht,lps},
employing the fact that $p$-brane solutions may be ``stacked up'' owing to
their zero-force properties. Both procedures preserve the $\Delta$ values
(\ref{avalue}) of $p$-brane solutions \cite{lpss,lps}. Then combining these
different reduction/oxidation procedures with duality transformations
permits relations to be established between different $p$-branes. For
example, combining a straight reduction from $D+10$ to $D=6$, a duality
transformation in $D=6$ and a diagonal oxidation from $D=6$ back to $D=10$
relates the $D=10$ string and 5-brane solutions. Details of these
applications will be given in \cite{lps}. 

\section*{Acknowledgements}

     We are grateful to M.J.\ Duff, C.M.\ Hull, E.\ Kiritsis, S.\ Mukherji, 
G.\ Papadopoulos, J.\ Rahmfeld, P.K.\ Townsend and A.\ Tseytlin
for helpful discussions.

\end{document}